\documentclass[prx,reprint,superscriptaddress,longbibliography]{revtex4-2}
\usepackage[colorlinks=true,allcolors=blue,breaklinks=true]{hyperref}
\usepackage{graphicx}
\usepackage{siunitx}
\usepackage{amsmath}

\usepackage[T1]{fontenc}
\usepackage{lmodern}
\usepackage{bbding}
\usepackage{xcolor}

\begin{document}
\title{Superfluid $^3$He aerogel experiments as a laboratory neutron star analogue}

% Authors
\author{S.~Autti}
\email{s.autti@lancaster.ac.uk}
\affiliation{Department of Physics, Lancaster University, Lancaster, LA1 4YB, UK.}

\author{V.~Graber}
\affiliation{Department of Physics, Royal Holloway University of London, Egham, Surrey, TW20 0EX, UK}

\author{B.~Haskell}
\affiliation{Dipartimento di Fisica, Universit\`{a} degli Studi di Milano, Via Celoria 16, 20133 Milano, Italy}
\affiliation{INFN, Sezione di Milano, Via Celoria 16, 20133 Milano, Italy}

%\date{\today}

%%%%%%%%%%%%%%%%%%%%%%%%%%%%%%%%%%%%%%%%%%%%%%%%%%%%%%%%%%%%%%%%%%%%%%%%%%%%%%
%%%%%%%%%%%%%%%%%%%%%%%%%%%%%%%%%%%%%%%%%%%%%%%%%%%%%%%%%%%%%%%%%%%%%%%%%%%%%%

%PRX format guidelines: main text no length limit, abstract about 5% of the main text but no more than 500 words

\begin{abstract} 
Neutron stars make a unique astrophysical test bench for our understanding of quantum physics at kilometre scales. The rotation of a neutron star features glitches, sudden spin-ups that interrupt the otherwise regular stellar spin-down, which are often attributed to the dynamics of pinned quantised vortices in one or several of the superfluid phases inside the star. Laboratory experiments probing superfluid vortices have inspired neutron star theory and simulations from the beginning. Here we argue that vortex experiments in superfluids contained in aerogels show phenomenology that offers a highly appealing but vastly unexplored analogue for neutron star physics. We build a point-vortex simulation that allows analysing experiments in a crust-like and a core-like aerogel, extracting two different regimes of pinned vortex (non-)dynamics and validating a microscopic picture of very strong vortex pinning. In the crust-like aerogel, vortices get depinned once the ambient superflow is fast enough, while in the core-like aerogel pinned vortices are never released and rotational velocity changes are accommodated by the avalanche-like production of new vortices instead. Finally, we show that these concepts should apply also in neutron stars and may thus revolutionise the analysis of neutron star observations.
\end{abstract}

\maketitle

In a superfluid far below the superfluid transition temperature, rotation is enabled by topological defects, typically line-like quantum vortices. The angular momentum of the superfluid can only change if the vortices move, if they are annihilated or if new vortices are created. Neutron stars are believed to consist of different phases of superfluid neutrons where vortices are pinned in place by underlying non-superfluid structures \cite{Sauls1989}. In the outer layers of a neutron star, superfluid neutron vortices are expected to pin to the nuclear clusters that form the crystalline crust. In the core of the star, pinning is thought to occur between superfluid neutron vortices and superconducting flux tubes, as protons are also expected to pair at high densities, and form a type-II superconductor \cite{Baym1969, Wood2022}. Records of rotating pulsars that spin down slowly often feature different patterns of sudden increases in the rotation period called glitches, a key observational handle of cosmic superfluidity. These glitches, now observed in over a hundred radio pulsars \cite{Basu2022}, are speculated to result from collective changes in the vortex configuration (see \cite{antonopoulou2022pulsar} for a review), but comprehensive analyses or simulations have neither been presented for neutron stars nor for close laboratory analogue systems.

Superfluid $^3$He is a compelling laboratory platform for investigating vortex dynamics in neutron stars. It features several stable superfluid phases when immersed in aerogels \cite{dmitriev2020superfluid,alles1999evidence,halperin2019superfluid,pollanen2012new,dmitriev2019superfluid,scott2024planar,dmitriev2020superfluid_planar,PolarDmitriev} mimicking the neutron star vortex environments. These phases exhibit a number of different stable vortex types \cite{RevModPhys.59.533,HQVs_prl,PhysRevLett.79.5058,PhysRevLett.75.3320,PhysRevResearch.6.043112,PhysRevLett.53.584} for which the total vortex line length or the total counterflow in the system can be measured non-invasively and in-situ, allowing experimental access to global vortex physics. The aerogels consist of mostly empty space (typically $>90\%$) but also contain solid structures with features down to and below the coherence length. These solid structures provide a high-density network of pinning sites for vortices, similar to that in a neutron star. Finally, hundreds to thousands of vortices can be created and be allowed to interact in a typical experiment in three dimensions. As we show in this paper, experimental access to multiple different phases and multiple vortex types within these phases allows mapping the phenomenological categories of vortex dynamics, preference between which is determined by the competition of three critical velocities. This phenomenological landscape could make superfluid dynamics in a neutron star accessible in simulations, thus facilitating the analysis of the stars' internal physics based on astronomical observations.

Once a quantum vortex is pinned at a pinning site, it cannot move even if the flow environment changes, until a critical velocity is reached. Let us consider a cylindrical pinning site with diameter $d\lesssim\xi_0$, where $\xi_0$ is the zero temperature coherence length, aligned with the vortex. The simplest estimate of the maximum pinning force per unit length can be obtained from the condensation energy gain due to the defect replacing a part of the vortex core \cite{1984ApJ...276..325A},  
\begin{equation} \label{pinning_estimate2}
F_\mathrm{p, cond.}\approx \frac{\sigma}{4 \pi^2 } \frac{k_\mathrm{F}^2  \Delta r}{\xi}  
\end{equation}
where $k_\mathrm{F}$ is the Fermi momentum, $\Delta$ is the superfluid gap, $r=d/2$, and $\xi(T)$ is the coherence length and $T$ is temperature. Note that this approximation is only valid at low temperatures. However, in superfluid $^3$He more energy is stored in the order parameter distortion around the pinning site so that the pinning force per unit length becomes \cite{rainer1977small,Makinen2019} 
\begin{equation} \label{pinning_estimate}
    F_{p} \sim  \sigma \frac{k_\mathrm{F}^2  \Delta^2 r}{k_\mathrm{B} T_\mathrm{c}\xi}, 
\end{equation}
where $k_\mathrm{B}$ is the Boltzmann constant and $T_\mathrm{c}$ is the superfluid transition temperature. This estimate is independent of the details of the superfluid phase and order parameter \cite{rainer1977small} and it is up to two orders of magnitude larger in superfluid $^3$He than the bare effect of the core displacement of Eq. (\ref{pinning_estimate2}). Experimentally Eq. (\ref{pinning_estimate}) has been substantiated by the observed characteristic length scale in the Larkin-Imry-Ma state of (polar-distorted) $^3$He-A in an aerogel \cite{li2013superfluid,volovik2008larkin}, determined by the energy contained in the order parameter distribution around the aerogel structures. The filling factor $\sigma$ in both equations accounts for the fraction of the unit length of the vortex core being occupied by the pinning site. For a cylindrical pinning site aligned with the vortex $\sigma=1$. 

In aerogel experiments, the normal fluid is clamped by the aerogel and therefore co-rotates with the container and with the pinned vortices. Therefore, other flow-related forces acting on the pinned vortex vanish and the only force that can overcome the pinning force is the Magnus force. The Magnus force per unit length reads
\begin{equation} \label{Magnus_force}
    \vec{F}_\mathrm{m} = \vec{\kappa} \times \rho_s (\Vec{v}_\mathrm{L} - \Vec{v}_\mathrm{s})
\end{equation}
(we use the expression for the isotropic B phase for simplicity). Here $\Vec{v}_\mathrm{s}$ is the superfluid flow velocity, $\Vec{v}_\mathrm{L}$ is the velocity of the vortex line, $\kappa=|\vec{\kappa}|$ is the circulation quantum, and $\rho_s$ is the superfluid mass density. The balance between the Magnus force and the (maximum) pinning force defines a critical depinning superflow velocity $v_\mathrm{c,p}$. If the local superflow velocity exceeds $v_\mathrm{c,p}$, the depinned vortices move as determined by mutual friction parameters $\alpha$ and $\alpha'$ \cite{bevan1997vortex}, being pushed towards the centre or out of the system, depending on the polarity of the vortices and the local flow direction. This is the traditional picture also applied to explaining neutron star rotation glitches \cite{1975Natur.256...25A}.

%%%%%%%%%%%%%%%%%%%%%%%%%%%%%%%%%%%%%%%%%%%%%%%%%%%%%%%%%%%%%%%%%%%%%%%%%%%%%%

\begin{figure*}[t!]
\includegraphics[width=0.95\linewidth]{pinning_aerogel_v2.png}%
\caption{Superfluid $^3$He as a laboratory neutron star simulator: ({\bf a}) A neutron star is composed of ({\bf b}) an s-wave neutron superfluid with an underlying solid lattice in the crust and a p-wave neutron superfluid with an underlying array of flux tubes housed by superconducting protons in the outer core (middle). These environments are geometrically analogous to superfluid $^3$He immersed in ({\bf c}) a crust-like and ({\bf d}) a core-like aerogel (see Table~\ref{table:comparison} for a detailed geometrical comparison). The aerogels are illustrated schematically (left) and detailed in the adjacent SEM images (right) adapted from \cite{ahmad2023silica,zhelev2016observation}. The neutron star and the aerogels in the $^3$He experiments are rotated along the vertical axis along the plane of the figure as shown in panel (a).} \label{schematic_aerogels_vs_NS}
\end{figure*}

%%%%%%%%%%%%%%%%%%%%%%%%%%%%%%%%%%%%%%%%%%%%%%%%%%%%%%%%%%%%%%%%%%%%%%%%%%%%%%

Experiments show that a pinned vortex system can react to changes in the rotation velocity also by adding more vortices to the system. Fundamentally, there are two mechanisms: (i) Vortices can enter through the edge of the system with the critical velocity $v_\mathrm{c,e}$ or (ii) be nucleated in the bulk of the system with the critical velocity $v_\mathrm{c,n}$ \cite{ruutu1997intrinsic,VolovikBook}. As discussed in more detail below, the system reacts to changes in rotation velocity qualitatively differently if $v_\mathrm{c,e}> v_\mathrm{c,n}$ rather than $v_\mathrm{c,e}< v_\mathrm{c,n}$. 

Pinned vortex dynamics have been explored both in isotropic aerogels with pinning sites organised as disordered strands \cite{yamashita2005pinning} and in highly anisotropic aerogels with rod-like pinning sites aligned with the rotation axis \cite{HQVs_prl}. The former resemble the crust of a neutron star where the underlying lattice structure provides a high density of sub-coherence length pinning sites, while the latter is reminiscent of the core of a neutron star with flux tubes providing an array of sub-coherence-length-sized rod-like pinning sites (see Fig.~\ref{schematic_aerogels_vs_NS}). Neither experimental system has been convincingly simulated nor unambiguously analysed in the literature.  

In both aerogels, vortex pinning is observed to be strong, and many vortices remain fixed in place even after spin down to zero rotation. This is in contrast with the canonical weak-pinning scenario for glitches considered in the neutron star literature where pinning delays or slows down the outward vortex motion as the star spins down \cite{1975Natur.256...25A, 1980PThPS..69..376P, 1984ApJ...276..325A}. In the isotropic aerogel, we observe $v_\mathrm{c,n} > v_\mathrm{c,p} > v_\mathrm{c,e}$. The threshold for vortex formation at the boundary (and also for absorption, which is easier) is sufficiently low so that the dynamics are entirely controlled by the depinnning velocity. It controls how vortices move in or out of the container as the rotation velocity is increased or decreased. In the anisotropic aerogel, we instead have $v_\mathrm{c,n} < v_\mathrm{c,p}$ ($v_\mathrm{c,e}$ is not relevant). This is a qualitatively different scenario, not systematically considered in the neutron star context before \cite{HaskellAB2020}. 
%\bh{ Only case where it was explicitly mentioned I can think of is our paper with Danai and Barenghi - \cite{HaskellAB2020}, but for the crust. There are previous works on turbulence, but cannot remember them explicitly discussing vortex nucleation}. 
It leads to pinned vortex ``non-dynamics'', with the main process being the avalanche-like vortex creation in bursts within the bulk of the superfluid, albeit the underlying microscopic mechanism remaining unresolved. Once inserted, vortices seem to be removed only by annihilation with adjacent vortices of opposite polarity. 

In this Article, we summarise the different aerogel experiments and simulate their key aspects in a simple two-dimensional point vortex model to extract the pinning details and remove any interpretation ambiguity. Based on these results, we then suggest an experimentally-tested picture of pinned vortex dynamics in neutron stars and propose how to apply well-tested simulation techniques for investigating the consequences of vortex dynamics in an astrophysical context. In the following, we label the aerogels ``crust-like'' and ``core-like'' for simplicity based on geometrical and circumstantial similarities with the neutron star components, but note that the superfluids in the crust and the core are not homogeneous in reality and different parts of the star may fall into either qualitative category of vortex dynamics.

%%%%%%%%%%%%%%%%%%%%%%%%%%%%%%%%%%%%%%%%%%%%%%%%%%%%%%%%%%%%%%%%%%%%%%%%%%%%%%
%%%%%%%%%%%%%%%%%%%%%%%%%%%%%%%%%%%%%%%%%%%%%%%%%%%%%%%%%%%%%%%%%%%%%%%%%%%%%%

\section*{Crust-like aerogel}

In the simplest picture, the inner crust of a neutron star consists of a normal-conducting electron component and an s-wave paired superfluid of neutrons surrounding a solid lattice of neutron-rich nuclei \cite{2008LRR....11...10C}. The lattice provides an array of spherical pinning sites with diameter $d \sim 10$\,fm, smaller than the neutron superfluid coherence length, $\xi_0 \sim 10-100$\,fm. The lattice spacing varies in a comparable manner from a few tens to a hundred fm depending on the depth within the crust \cite{2018ApJ...865...23G}. The vortex spacing for a superfluid neutron star rotating at typical angular velocities in the range $\sim 0.5 - 500$\,rad/s is several orders of magnitude larger than any of these scales. Table \ref{table:comparison} lists the key parameter values in all systems considered in this Article.
% \vgc{[refer to the table if we add that?]}

The closest nanomaterial that has been studied in rotation with a structure resembling the neutron star's inner crust is a silica aerogel. The silica aerogel used in \cite{yamashita2005pinning} is composed of $98\%$ empty space with disordered, interconnected spherical silica particles ($d \approx 5$\,nm) forming disordered strands that make the aerogel (see Fig.~\ref{schematic_aerogels_vs_NS}). The typical spacing between the strands is $\sim 100$\,nm and the vortex spacing for achievable rotation velocities is of the order of \SI{50}{\micro\meter} or larger. Therefore, in units of the coherence length ($\xi_0 \approx 20$\,nm) the geometry closely resembles the neutron star crust. The sample container in \cite{yamashita2005pinning} is of cylindrical shape with radius $R=2.6$\,mm.

%%%%%%%%%%%%%%%%%%%%%%%%%%%%%%%%%%%%%%%%%%%%%%%%%%%%%%%%%%%%%%%%%%%%%%%%%%%%%%

\begin{figure}[t!]
\includegraphics[width=1\linewidth]{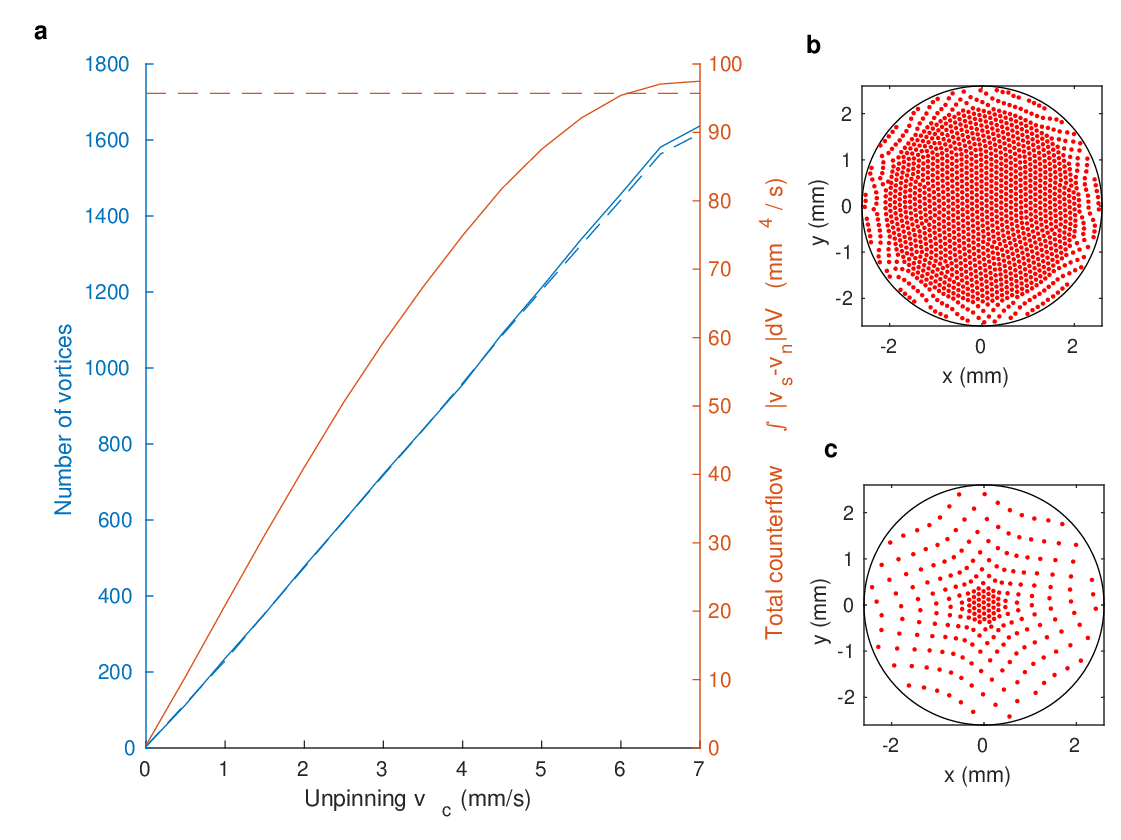}%
\caption{Spin-down simulation in the crust-like aerogel: ({\bf a}) The vortex pinning simulation starts from a steady-state configuration obtained with no pinning and $\Omega=2.6$\,rad/s. Rotation is then removed and the configuration allowed to equilibrate. The equilibrium vortex number increases linearly (solid blue line, left y axis) as a function of the depinning flow velocity $v_\mathrm{c}$. Starting from an initial condition with added 100\,$\mu$m positional noise (dashed blue line, left y axis) does not change the outcome. The experiment \cite{yamashita2005pinning} measures integrated counterflow (yellow line, right y axis), which reaches the solid-body value for $\Omega=2.6$\,rad/s (horizontal dashed line, right y axis) at around $v_c=6$\,mm/s. ({\bf b}) The final vortex configuration for $v_c=5.5$\,mm/s shows how the combined superflow of the vortices pushes the vortices outside of a certain distance from the centre outwards, and around 300 vortices leave the container. ({\bf c}) At smaller critical velocities a larger share of the vortices gets pushed out during the equilibration process (here $v_c=1$\,mm/s). The mutual friction parameters used in this simulation were $\alpha=2$ and $\alpha'=0.8$, corresponding to $T/T_\mathrm{c} \approx 0.8$ in the B phase at 30\,bar \cite{bevan1997vortex}, but we have confirmed that the outcome does not depend on the mutual friction parameters used. \label{FIG:cylinder_pinning_simulation}}
\end{figure}

%%%%%%%%%%%%%%%%%%%%%%%%%%%%%%%%%%%%%%%%%%%%%%%%%%%%%%%%%%%%%%%%%%%%%%%%%%%%%%

\begin{figure}[htb!]
\includegraphics[width=1\linewidth]{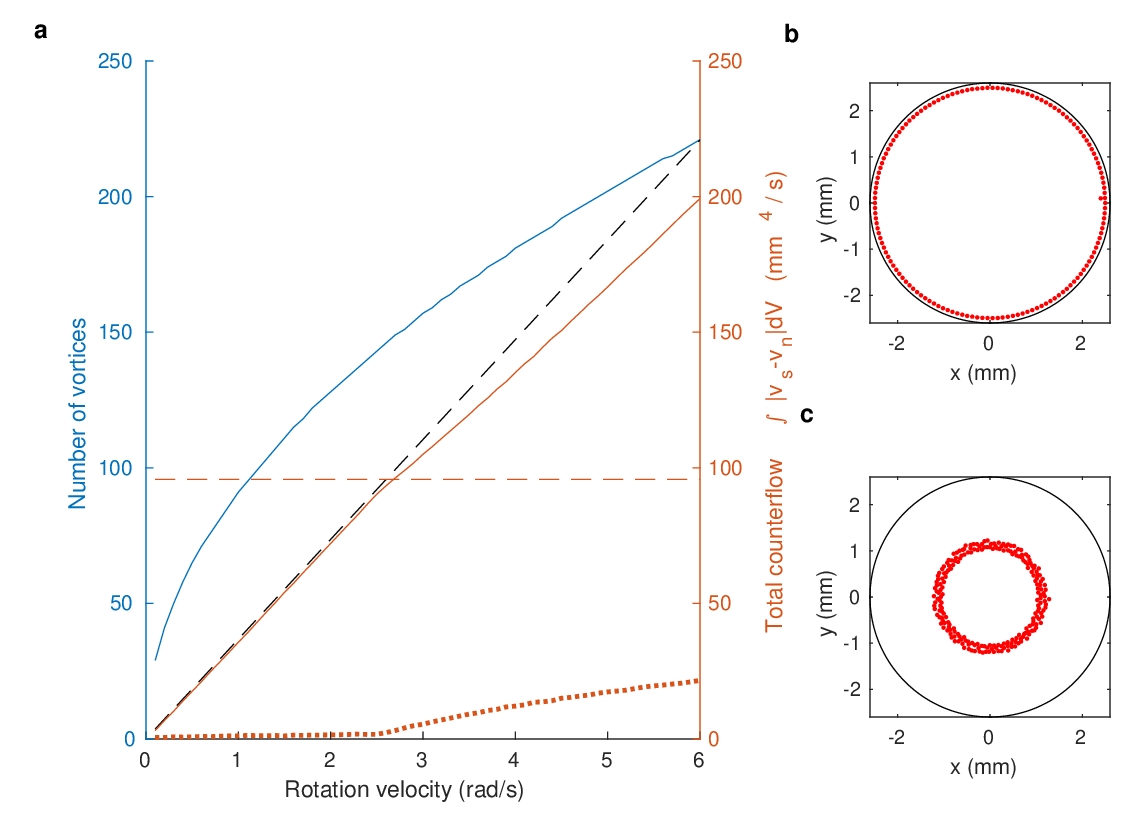}%
\caption{Spin-up simulation in the crust-like aerogel: ({\bf a})  To demonstrate how vortices enter the cell with increasing rotation velocity, we place a ring of vortices along the perimeter with an intervortex spacing that corresponds to each rotation velocity. We then let the system evolve until all vortices are pinned. The depinning superflow velocity is set to $v_\mathrm{c}=6$\,mm/s. Panel (a) shows how the first ring entering the system from the sliver of bulk B phase surrounding the aerogel would qualitatively affect the dynamics. The number of vortices grows with rotation velocity (blue line, left y axis), but the first ring has a negligible effect on the total integrated counterflow (yellow line, right y axis), because it adds very little to the integrated superflow (yellow dotted line, also shown on the right y axis), until the superflow velocity at the perimeter exceeds $v_\mathrm{c}$ at $\Omega_\mathrm{c}=2.6$\,rad/s. Above this rotation velocity the first ring moves inwards, increasing the total superflow and thus reducing the total counterflow as compared with the solid-body counterflow (black dashed line, right y axis). The reduction starts when the solid body counterflow reaches that corresponding to $\Omega=2.6$\,rad/s (horizontal dashed line, right y axis). Allowing for consecutive rings to enter would further enhance this effect and allow the counterflow to saturate near the solid-body counterflow at $\Omega=2.6$\,rad/s, as observed in the experiment. ({\bf b}) The vortex configuration for $\Omega=2.5$\,rad/s is below the threshold and thus does not move further into the aerogel from the initial position, chosen to be 100\,$\mu$m inwards from the outer wall to approximate the thickness of the bulk phase around the aerogel \cite{bunkov2000semisuperfluidity}. ({\bf c}) At $\Omega=6$\,rad/s the first ring entering the system from the perimeter ends up at a 1.2\,mm distance from the centre where the superflow velocity in the absence of vortices is equal to $v_\mathrm{c}$. The mutual friction parameters used in this simulation were $\alpha=2$ and $\alpha'=0.8$, corresponding to $T/T_\mathrm{c} \approx 0.8$ in the B phase at 30\,bar \cite{bevan1997vortex}, but we have confirmed that the outcome does not depend on the mutual friction parameters used. \label{FIG:spinup_simulation}}
\end{figure}

%%%%%%%%%%%%%%%%%%%%%%%%%%%%%%%%%%%%%%%%%%%%%%%%%%%%%%%%%%%%%%%%%%%%%%%%%%%%%%

\subsection*{Experimental observations}

The superfluid $^3$He placed within the crust-like silica aerogel is in the B phase \cite{alles1999evidence}, which for the purposes of vortex mechanics is as close to an s-wave system as one can get in superfluid $^3$He; the B phase gap spectrum is (nearly) isotropic. Thus, the pinning force and the Magnus force are direction-independent as in Eqs.~(\ref{pinning_estimate},\ref{Magnus_force}). The vortex configuration in this system can be inferred from a nuclear magnetic resonance (NMR) measurement of total counterflow $\int \mathrm{d}V |\Vec{v}_\mathrm{n} - \Vec{v}_\mathrm{s}|$ in the sample. That is, in the laboratory frame the superflow field of vortices pinned in place creates $\Vec{v}_\mathrm{s}$, while the normal fluid rotates with the container because it is locked to the aerogel, creating $\Vec{v}_\mathrm{n}$. The total integrated counterflow in the sample can be inferred from the NMR line shape measured for the entire superfluid volume in the aerogel \cite{yamashita2005pinning}.

The authors in \cite{yamashita2005pinning} cool the sample to the superfluid state in zero rotation at 30\,bar pressure. When the rotation velocity $\Omega$ is spun up from zero, the measured counterflow increases linearly until $\Omega_\mathrm{c} \approx 2.6$\,rad/s is reached. Up to this velocity no new vortices are created, meaning that the superfluid component does not rotate while the normal component's angular velocity increases. At $\Omega_\mathrm{c}$ the proportionality breaks, marking the creation of vorticity which reduces the increase in counterflow compared to the linear behaviour. This threshold corresponds to a critical flow velocity of $v_\mathrm{c}=6.7$\,mm/s at the container wall where the solid-body flow is the largest and, in a naive picture, vortices are created. The rotation velocity is then increased further and eventually decreased back to zero. While the normal component is spun down completely as a result, the counterflow does not vanish. Instead, the counterflow that is locked in by vortex pinning at $\Omega=0$ corresponds to about that at $\Omega_\mathrm{c} \approx 2.6$\,rad/s before vortices were inserted. At this counterflow velocity (but not below), the mismatch between the superflow and the normal flow is sufficiently large to depin vortices where the counterflow is fastest. 

\subsection*{Simulations}

We can recreate these observations in a numerical model of the system. We simulate the system in a two-dimensional point vortex model where vortices are only allowed to move if the Magnus force acting on them exceeds the pinning threshold that corresponds to the local flow velocity $v_\mathrm{c,p}$. We will return to deriving this depinning velocity below, but treat it as a fitting parameter for now. Pinning is approximated to be continuously available everywhere. This is justified later.

Let us first study the now-stationary container in the spin-down experiment that needs to hold a sufficient number of vortices pinned in place to lock in the observed counterflow. The equilibrium number of vortices in this container at 2.6\,rad/s rotation is $N\approx1600$. The configuration that minimises the forces between the vortices is the dynamic equilibrium state at this rotation velocity in the absence of pinning. This is used as the starting point of the simulation. The configuration is then allowed to evolve with pinning turned on and rotation turned off until all vortices are either pinned in place or have moved out of the container, and the configuration is thus stable. 

The final stable vortex number is shown in Fig.~\ref{FIG:cylinder_pinning_simulation} as a function of $v_\mathrm{c,p}$. We find that $v_\mathrm{c,p}\approx 6$\,mm/s is needed to freeze in a total counterflow equal to the solid body rotation at $\Omega= 2.6$\,rad/s. The final vortex configuration contains $N\approx1400$ vortices, which is somewhat smaller than the dynamic steady state value at $\Omega= 2.6$\,rad/s without pinning, because vortices near the perimeter freeze in last, but they contribute almost nothing to the integrated superflow. We emphasise that the vortex configuration in the experiment may not be distributed as shown in Fig.~\ref{FIG:cylinder_pinning_simulation}b,c due to, for example, the initial state used in the simulation not replicating that resulting from spin-up and spin-down. That is, the below spin-up simulations hint that the vortices entering the system from the perimeter may not populate the very centre as densely as the rest of the system. In this case, even stronger pinning would be required to lock in the observed counterflow. The role of vortices with opposite polarity is discussed later.

Second, we need to explain how vortices enter the system in the spin-up experiments when the rotation velocity is increased from zero, initially in the absence of vortices. In this system, the B phase within the aerogel is surrounded by a sliver of bulk B phase in the gap between the aerogel and the container wall. We expect that vortices can easily enter and exit the system through the edge of the aerogel due to the absence of a phase interface in this area.
Outside the aerogel, at the rough walls of the sample container, vortex nucleation typically takes place at sub mm/s velocities (with specifically-polished smooth walls this is an order of magnitude larger, roughly one mm/s \cite{ruutu1997intrinsic,PhysRevLett.99.265301,eltsov2024kelvin,skrbek2003vortex,eltsov2019kelvin}),
effectively resulting in $v_\mathrm{c,e} \approx 0$. However, the experiment is not sensitive to the precise value as long as $v_\mathrm{c,e}<v_\mathrm{c,p}$. Consequently, the key feature stopping vortices from moving inwards is pinning by the aerogel. A point vortex calculation offers no self-consistent ways of allowing for such vortex nucleation. Instead, we place a ring of pinned vortices along the perimeter of the sample container with an intervortex spacing that corresponds to the uniform equilibrium state at each rotation velocity. The vortices are placed 100\,$\mu$m from the wall to allow for the thickness of the layer of bulk superfluid \cite{alles1999evidence,yamashita2005pinning}. We then let the system evolve until all vortices are again pinned in place. The depinning superflow velocity is taken from the spin-down simulation above ($v_\mathrm{c}=6$\,mm/s) with the intention of showing that the experiment is exhaustively characterised by this one parameter value.  

With these assumptions, the simulated number of vortices in the container grows with the rotation velocity (Fig.~\ref{FIG:spinup_simulation}). However, these vortices remain pinned in place and, thus, have a negligible effect on the total integrated superflow, and counterflow, up to $\Omega_\mathrm{c}=2.6$\,rad/s. Above $\Omega_\mathrm{c}=2.6$\,rad/s, the first ring moves inwards, increasing the total superflow and thus reducing the total counterflow, as observed in the experiment. Note that the outer wall attracts the pinned vortices (due to opposite-polarity image vortices on the other side of the wall), and the vortex motion is triggered at a marginally larger rotation velocity than that corresponding to 6\,mm/s solid-body flow at the wall (2.6\,rad/s vs 2.3\,rad/s rotation).  Allowing for consecutive rings to enter would further enhance the counterflow reduction, but this cannot be done self-consistently in the present model. Based on the simulation results for the spin-up and spin-down experiments, we conclude that in the aerogel in \cite{yamashita2005pinning} the depinning velocity is $v_\mathrm{c,p} \approx 6$\,mm/s. 

We now turn our attention to $v_\mathrm{c,n}$. Motivated by the polar phase experiments discussed in the next section, let us assume for a moment that $v_\mathrm{c,n}<v_\mathrm{c,p}$. During a spin-up in a vortex-free state, the first vortices would enter the system when the local flow velocity near the container wall reaches $v_\mathrm{c,n}\approx6.7$\,mm/s at $\Omega_\mathrm{c}=2.6$\,rad/s. Thus, during the spin-up the integrated counterflow would grow linearly up to $\Omega_\mathrm{c}=2.6$\,rad/s, as observed in \cite{yamashita2005pinning}. Above this velocity the counterflow would decrease abruptly with an avalanche of vortices created, as observed in the experiment discussed in the next section \cite{PhysRevResearch.2.033013}. This would not result in a smooth levelling off of the total counterflow as observed in \cite{yamashita2005pinning}. Finally, during the spin down to $\Omega=0$, the system would add new vortices with opposite polarity (some of which get pinned in place while others annihilate existing vortices) to accommodate the changes in rotation instead of removing existing vortices. Locking in the counterflow that corresponds to solid body rotation at $\Omega=2.6$\,rad/s does not naturally follow from $v_\mathrm{c,n}\approx 6.7$\,mm/s as it does if the changes in counterflow result from vortex motion. That is, with an existing array of vortices pinned in place in the aerogel, the number of vortices taking part in the spin-down depends on the details of the local flow patterns that result from the avalanches. While this may, in principle, result in any number of opposite polarity vortices being created, a specific number such as that corresponding to the maximum counterflow the system can contain during spin-up requires fine tuning the details of the process. In addition, repeated spin-up and spin-down cycles always result in the same counterflow at $\Omega=0$ in the experiment \cite{yamashita2005pinning}. While this behaviour is naturally explained in the simulations outlined above (because the maximum counterflow the system is able to contain at $\Omega=0$ only depends on the available pinning force to first approximation and not on the details of the vortex configuration), such behaviour is difficult to reconcile with the vortex-antivortex accumulation picture. In contrast, in the experiment in the next section the total vortex number keeps increasing during cyclic spin-up and spin-down conditions and it is difficult to see how this ever increasing density of pinned vortices would not change the patterns of vortex nucleation in the bulk of the system, thus also changing the counterflow observed at $\Omega=0$ after repeated cycles. We therefore conclude that $v_\mathrm{c,n}<v_\mathrm{c,p}$ is likely inconsistent with the experimental observations in \cite{yamashita2005pinning}.

%%%%%%%%%%%%%%%%%%%%%%%%%%%%%%%%%%%%%%%%%%%%%%%%%%%%%%%%%%%%%%%%%%%%%%%%%%%%%%

\begin{figure}[htb!]
\includegraphics[width=1\linewidth]{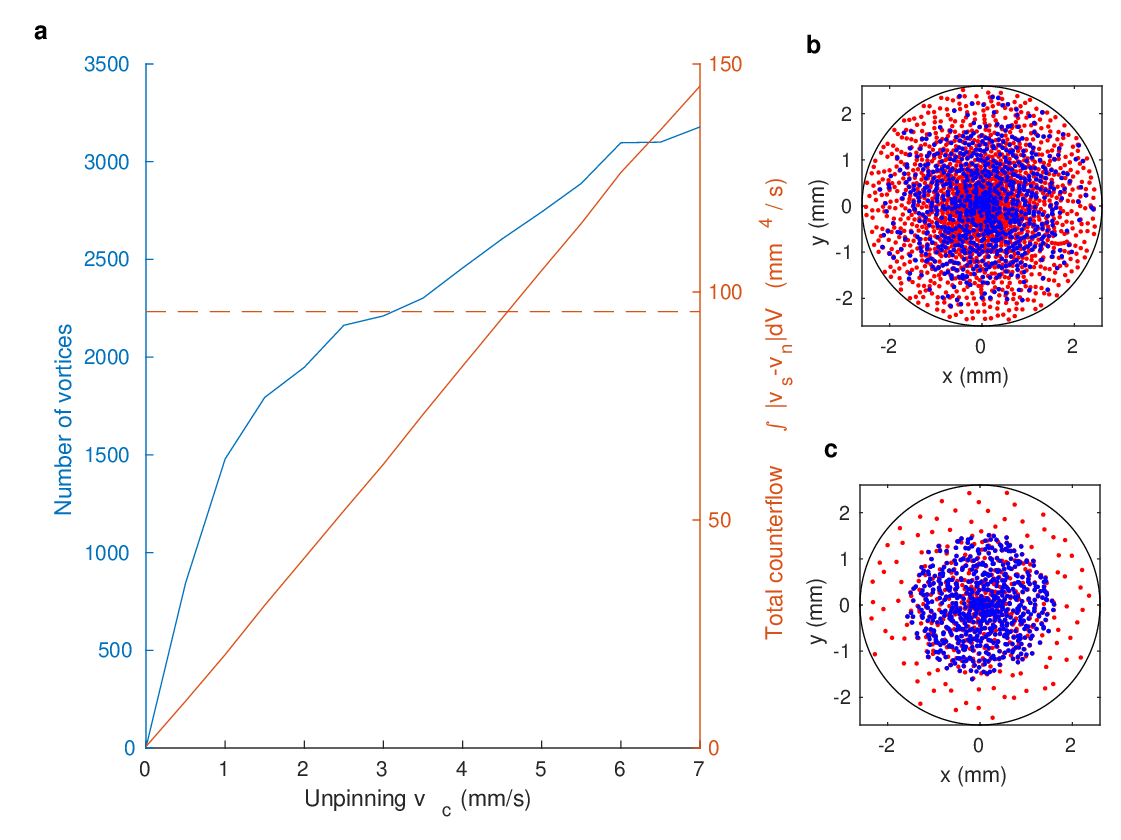}%
\caption{Spin-down simulation with negative and positive polarity vortices in the crust-like aerogel: ({\bf a}) Repeating the simulation shown in Fig.~\ref{FIG:cylinder_pinning_simulation} from a starting configuration with twice the initial vortex number (3200), with 50\% of the extra vortices added having the opposite polarity, does not change the previously observed phenomenology. However, the number of vortices that remain pinned for a given $v_\mathrm{c}$ is increased compared to Fig.~\ref{FIG:cylinder_pinning_simulation} as the negative polarity vortices reduce the total flow in the container. Hence, the threshold for locking in the total counterflow that corresponds to solid body rotation at $\Omega = 2.6$\,rad/s (horizontal dashed line) is reduced by 1.5\,mm/s to $v_\mathrm{c} \approx 4.5$\,mm/s. ({\bf b}) The final state at $v_\mathrm{c} = 5.5$\,mm/s consists of a noisy array of positive polarity vortices (red dots) whose density decreases towards the boundary, and a similar distribution of negative polarity vortices (blue dots) whose total number is smaller, reflecting the vortex population and positive total counterflow of the initial state. ({\bf c}) At a smaller depinning velocity, $v_\mathrm{c} \approx 1$\,mm/s, the density of vortices in the final state is lower, but the general pattern remains similar. We note that the final distribution depends on the noise added to the initial state (here uniform positional noise with maximum deviation of 100\,$\mu$m) and the total initial vortex number, but the qualitative observations are independent of these details. \label{FIG:cylinder_pinning_simulation2}}
\end{figure}

%%%%%%%%%%%%%%%%%%%%%%%%%%%%%%%%%%%%%%%%%%%%%%%%%%%%%%%%%%%%%%%%%%%%%%%%%%%%%%

It is also possible that $v_\mathrm{c,n} \approx v_\mathrm{c,p}$. To simulate this possibility, we start with an equilibrium array of vortices at $\Omega=2.6$\,rad/s with 1600 vortices, adding a randomly positioned population of 800 positive and 800 negative polarity vortices to the system. In this configuration a larger number of vortices is fixed in place by a given $v_\mathrm{c,p}$ at zero rotation (Fig.~\ref{FIG:cylinder_pinning_simulation2}) as compared with Fig.~\ref{FIG:cylinder_pinning_simulation}. This is because the flow fields of the positive and negative vortices largely cancel out. Importantly, the screening provided by the negative vortices also means that the critical velocity needed to freeze in the counterflow that corresponds to solid-body rotation at $\Omega=2.6$\,rad/s is reduced to $v_\mathrm{c,p}\approx 4.5$\,mm/s. That is, negative vortices help the system lock in more total counterflow. Such a reduction is inconsistent with what was observed by \cite{yamashita2005pinning} because the critical velocity during spin-up was also determined by $v_\mathrm{c,p}$. Thus, negative polarity vortices may be created during the spin-down (or even spin-up), but these constituting a sizeable fraction of the total vortex number is inconsistent with the experimental observations. Putting these and the above observations together, we conclude that $v_\mathrm{c,e}<v_\mathrm{c,p}<v_\mathrm{c,n}$ and $v_\mathrm{c,p} \approx 6$\,mm/s.

\begin{figure}[htb!]
\includegraphics[width=1\linewidth]{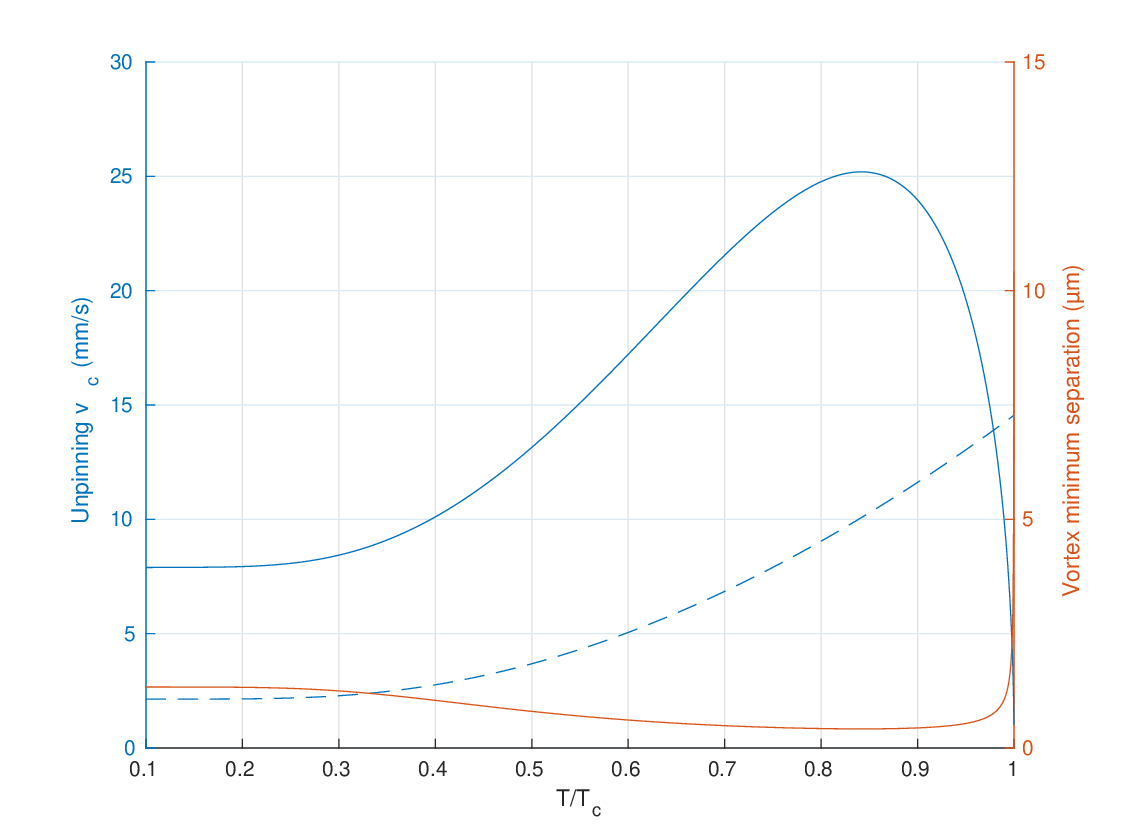}%
\caption{Vortex pinning in the crust-like aerogel: The vortex depinning superflow velocity is estimated by balancing Eqs.~(\ref{pinning_estimate},\ref{Magnus_force}) with $d=5$\,nm for a cylindrical defect with filling factor $\sigma=1/20$, approximating the structure of the crust-like aerogel (blue line, left y axis). At low temperatures this estimate is in good agreement with the velocity extracted from the experiment using the simulations ($v_\mathrm{c}= 6$\,mm/s) but it only coarsely captures the experimentally-observed temperature independence between $T/T_\mathrm{c}=0.47$ and $T/T_\mathrm{c}=0.83$ \cite{yamashita2005pinning}. Importantly, the unpinning velocity estimate goes to zero at the superfluid transition temperature, thus allowing for an equilibrium array of vortices to form if the system is cooled down slowly in rotation. This is a key assumption made in interpreting the experiment. The unpinning velocity resulting from the bare core displacement effect obtained by equating Eqs.~(\ref{pinning_estimate2},\ref{Magnus_force}) (dashed line, left y axis) is drawn here for comparison. We set $\sigma=1$ for this estimate, because using a more realistic filling factor would make the values too small to see in this plot. We also calculate the minimum stable distance between two otherwise isolated vortices (yellow line, right y axis) by finding the distance at which the superflow created by one vortex reaches $v_\mathrm{c,p}$ given by the solid line on the left y axis at the location of the other vortex. The parameter values and expressions used here correspond to the temperature-dependent bulk B phase $\Delta$, $\xi$ and $\rho_\mathrm{s}$ at 30\,bar pressure, which may not perfectly reflect those in the aerogel but provide a reasonable estimate of the corresponding dynamics. \label{FIG:pinning_forces}}
\end{figure}

The strongest argument to support this result we can construct is deriving $v_\mathrm{c,p}$ from theory. We solve for $v_\mathrm{c,p}$ by setting $F_\mathrm{p}=F_\mathrm{m}(v_\mathrm{c,p})$ (Fig.~\ref{FIG:pinning_forces}). To obtain a realistic estimate of the depinning velocity, we need to account for the chains of spherical pinning sites not being aligned with vortices in most cases. Given the inter-strand distance and pinning site size quoted above, the filling factor is $\sigma \sim 1/20$. This filling factor is justified, because at 100\,nm scales (comparable with the coherence length) the vortices are approximately straight and thus mostly misaligned with the silica strands, as a significant vortex bending radius at this length scale would require a forcing flow of the order of the Landau velocity $v_\mathrm{L}\approx 70$mm/s (at 30\,bar) \cite{VolovikBook}. Further, owing to the large pinning energy, mesoscopic spatial averaging of the pinning force as a function of vortex orientation as discussed in \cite{Mesoscopic16} is also insignificant (this is elaborated in the NS section). The depinning critical velocity obtained from these considerations is most reliable at low temperatures, where the estimate yields $v_\mathrm{c,p}\sim8$\,mm/s in good agreement with the $v_\mathrm{c,p}\approx6$\,mm/s extracted from the experiment \cite{yamashita2005pinning} using our simulations above. The experiment probes the temperature rang between $T/T_\mathrm{c}=0.47$ and $T/T_\mathrm{c}=0.83$ \cite{yamashita2005pinning}, where the critical velocity is reported to be temperature independent. The theoretical pinning force estimate does not quite capture this independence. We, however, note that the pinned vortices overlap with several aerogel strands at higher temperatures (the core size is $\sim\xi$, which increases with temperature) and the resulting averaging \cite{Mesoscopic16} may reduce the pinning force as temperature increases, thus acting to level off the maximum seen in Fig.~\ref{FIG:pinning_forces} at $T=0.85 T_\mathrm{c}$. For these reasons the estimated $v_\mathrm{c,p}$ close to $T_\mathrm{c}$ should be taken as indicative only. We conclude that (i) the pinning force of Eq.~(\ref{pinning_estimate}) arising from the order parameter distortion around the pinning sites matches experimental observations without the need for fitting additional parameters at low temperatures, that (ii) it provides indicative agreement at higher temperatures, and that (iii) the much smaller core displacement effect Eq.~(\ref{pinning_estimate2}) is orders of magnitude too small to account for the observed pinning, as shown in Fig.~\ref{FIG:pinning_forces}.

Converting the resulting $v_\mathrm{c,p}$ to the smallest stable distance between two vortices (Fig.~\ref{FIG:pinning_forces}) yields $l\sim $\SI{1}{\micro\meter} up to $T/T_\mathrm{c}=0.99$, above which the distance quickly reaches a fraction of a mm. Below this distance vortices of the same polarity push another further apart and opposite polarity pairs are free to annihilate one another. Even the low temperature distance is much larger than the typical spacing of physical features in the aerogel (or the vortex cores), thus substantiating the assumption made above that pinning sites are effectively continuously available everywhere. This minimum distance also sets a lower bound for the spatial resolution needed in large vortex simulations, which can potentially be used to reduce the computational difficulty of large-scale neutron star vortex dynamics simulations.

%%%%%%%%%%%%%%%%%%%%%%%%%%%%%%%%%%%%%%%%%%%%%%%%%%%%%%%%%%%%%%%%%%%%%%%%%%%%%%

%%%%%%%%%%%%%%%%%%%%%%%%%%%%%%%%%%%%%%%%%%%%%%%%%%%%%%%%%%%%%%%%%%%%%%%%%%%%%%
%%%%%%%%%%%%%%%%%%%%%%%%%%%%%%%%%%%%%%%%%%%%%%%%%%%%%%%%%%%%%%%%%%%%%%%%%%%%%%

% conclude that the model works?

\section*{Core-like aerogel}

The core of a neutron star consists of a p-wave paired superfluid of neutrons, an s-wave paired proton superconductor state, an unpaired electron/muon component and potentially exotic particles at the highest densities \cite{1989ASIC..262..457S, 2017IJMPD..2630015G, 2017JApA...38...43C}. In the simplest picture, the neutron star's large magnetic field is confined to an array of aligned cylindrical flux tubes in the superconductor, with diameter $\sim$100\,fm (similar in order to the neutron coherence length) and a spacing that is an order of magnitude larger for canonical field strengths of $\sim 10^{12}$\,G. These flux tubes provide a high density of cylindrical pinning sites for vortices in the neutron superfluid, with the typical neutron superfluid vortex separation being several orders of magnitude larger than the pinning site spacing.

The closest nanomaterial with a structure resembling this arrangement is Nafen aerogel. Nafen aerogel is formed of an array of parallel solid strands with diameters $d\sim10$\,nm (see Fig.~\ref{schematic_aerogels_vs_NS}). The typical spacing of these strands is $\sim$50\,nm, but several Nafen densities can be created and have been studied in superfluid experiments \cite{Asadchikov2015}. The superfluid $^3$He coherence length is $\sim50$\,nm, which means that geometrically Nafen aerogel closely mimics the core of a neutron star, and we label it ``core-like''. As for the crust-like aerogel, the inter-vortex spacing in practical experiments in the core-like aerogel is typically $\sim$100\,$\mu$m, and thus the pinning site density is essentially continuous for mesoscopic purposes. 

\subsection*{Experimental observations}

Placing superfluid $^3$He within Nafen aerogel modifies the superfluid phases available. Most of the vortex experiments have been carried out in the polar phase \cite{PolarDmitriev,Mineev2016,zhelev2016observation} in a cubic container \cite{HQVs_prl}. In the polar phase, there are two types of phase vortices associated with mass flow: half-quantum vortices \cite{HQVs_prl} (HQV) that carry a half-quantum of mass circulation and a half-quantum of spin circulation, and usual single-quantum phase vortices \cite{PhysRevResearch.2.033013} (SQV). The total line length of HQVs can be inferred experimentally from the continuous-wave NMR line satellite feature they produce \cite{HQVs_prl,PhysRevLett.127.115702}, by integrating the area under the satellite. In contrast, the total line length of SQVs can be inferred from the relaxation rate $1/\tau$ of a magnon quasiparticle condensate created in the same volume \cite{PhysRevResearch.2.033013,autti2018bose}. Note that while the total line length of each vortex type (absolute value) can be accessed separately, the vortex polarities or the counterflow cannot be measured directly. The applied magnetic field direction and strength determine whether SQVs or HQVs are energetically favourable and thus which type is created if the sample is slowly cooled down to the superfluid state in rotation. 

There are several open problems in the interpretation of the core-like aerogel experiments that can be directly or indirectly studied using the pinned point vortex model developed in the previous section. Addressing these issues removes ambiguity of interpreting the phenomenology of the observed vortex non-dynamics, while simultaneously substantiating the simulation techniques used here. This approach also lays down criteria for applying the observations to other systems, such as neutron stars.

The authors of \cite{HQVs_prl} report that cooling down in steady rotation and a magnetic field orientation that favours HQVs results in an array of HQVs with total vortex line length (i.e., number of vortices for rectilinear vortices) $\propto \Omega$, as expected. In contrast, the observed relaxation rate when SQVs are created by cooling down in rotation, and no HQVs are present (as confirmed by the absence of the NMR satellite line), is $1/\tau(\Omega) - 1/\tau(0) \propto \sqrt{\Omega}$ \cite{PhysRevResearch.2.033013}. The SQVs are not associated with any structures in the order parameter distribution that has this scaling. Moreover, based on similar bulk B phase experiments \cite{autti2021vortex}, we would expect the change in the relaxation rate to be proportional to the vortex number (total line length in the sample) and thus $\propto \Omega$. This unexpected scaling is the first question we address below using the simulations.

A rotation-independent channel of HQV production has also been identified \cite{PhysRevResearch.2.033013}: When passing through the second-order phase transition at $T_\mathrm{c}$, the fluctuation-driven Kibble-Zurek (KZ) mechanism produces a randomly-oriented tangle of all available types of topological defects, including HQVs (unless the KZ mechanism is suppressed by a symmetry violating bias \cite{PhysRevLett.127.115702}). Experimental observations suggest that the KZ vortices in Nafen aerogel do not decay rapidly, as they do in bulk superfluid experiments \cite{KZ_nature,KZ_nature2,PhysRevB.90.024508,ProgLowTempPhys_page9}. Instead, the full density of initially produced vortices is pinned in place by the aerogel. This is the second aspect we will study numerically.

Finally, the most relevant experiments in the core-like aerogel start with cooling to the superfluid state in zero rotation \cite{PhysRevResearch.2.033013}. The rotation velocity is then increased to $\Omega=2$\,rad/s and subsequently decreased back to zero. The system responds to the increase in the flow by producing SQVs in sudden avalanches with a critical velocity $v_\mathrm{c}\approx 2$\,mm/s regardless of which vortex type is made energetically favourable by changing the orientation and magnitude of the applied magnetic field. (This behaviour reflects the findings in $^3$He-A \cite{blaauwgeers2000double,rantanen2025structure}, where despite never being energetically favoured the double quantum vortex has a lower critical velocity than the single-quantum vortex and is thus created in similar low-temperature spin-up experiments.) During spin-down, the vortex number increases further. We use the understanding of depinning physics obtained previously to argue that this process is related to $v_\mathrm{c,n}$ and thus cannot be described within a point vortex model. The experiments reported in \cite{PhysRevResearch.2.033013} also show that applying as rapid a change in the rotation velocity as permitted by the equipment additionally increases the HQV number, presumably by driving the SQV system far out of equilibrium and thus allowing for flow velocities beyond the SQV threshold at $v_\mathrm{c}\approx 2$\,mm/s. However, the precise onset of this process was not studied carefully. We will also investigate this observation in light of our vortex pinning picture below.

We would also like to emphasise that \cite{HQVs_prl} reports that the HQVs are pinned in place strongly enough so that their number does not change at any temperature below $T_\mathrm{c}$. This holds even if the rotation applied to create the vortices is brought to zero and the temperature increased to above $0.95 \, T_\mathrm{c}$ for hours. Strong pinning is further supported by the observation that, in a lower-density version of the same aerogel with the B phase stable at low temperatures, the HQVs can be taken from the polar phase to the B phase by decreasing the temperature. In the B phase, the HQVs are unstable and pulled together by a domain wall that connects adjacent HQVs. Yet, returning to the polar phase revealed no changes in the HQV number or positions \cite{Makinen2019}. 

\subsection*{Simulations}

The experimental container in \cite{HQVs_prl,PhysRevResearch.2.033013,PhysRevLett.127.115702} is a 4\,mm$\times$4\,mm$\times$4\,mm box, filled with the aerogel. We model it as a square, thus neglecting possible disorder in the z dimension from, say, pinned KZ vortices \cite{HQVs_prl,PhysRevLett.127.115702}. The pure potential flow (that is, superflow in a rotating container in the absence of vortices) in the box is not the trivial solid body rotation that applies in a cylinder. In a box, the superflow is enhanced near the middle points of each wall. We obtain this flow pattern by solving the Laplace equation numerically for this geometry \cite{PhysRevResearch.2.033013} and then feed the solution into the point vortex simulation. 

To construct the point vortex simulations, we need an order-of-magnitude understanding of the depinning velocity $v_\mathrm{c,p}$. We first note that the pinning sites in the core-like aerogel are cylindrical and aligned with the rotation axis (see Fig.~\ref{schematic_aerogels_vs_NS}). Therefore, for vortices created by rotating the sample, the filling factor is $\sigma \approx 1$. This means that the expected $v_\mathrm{c,p}$ is an order of magnitude (even two) larger than in the crust-like aerogel. There are also two less important differences: Unlike in the B phase, the polar phase gap spectrum is not homogeneous due to the presence of a node line where the gap is zero in momentum space \cite{PolarDmitriev, Mineev2016,vollhardt2013superfluid}. Thus, a precise description of the Nafen superfluid dynamics would require modifying Eqs.~(\ref{pinning_estimate},\ref{Magnus_force}) qualitatively in a similar fashion as in the A phase \cite{bevan1997vortex}: The superfluid gap suppression in the plane of the gap node line in principle decreases both the pinning force and the Magnus force. However, Eq.~(\ref{pinning_estimate}) has been substantiated experimentally in the polar-distorted A phase in the same aerogel type \cite{Makinen2019,volovik1996glass,li2013superfluid,dmitriev2010orbital} and can thus be expected to work also in the polar phase. On the other hand, the Magnus force is reduced by the decrease of the superfluid density in the plane of the gap suppression \cite{bevan1997vortex}, which acts to increase $v_\mathrm{c,p}$, strengthening the expectation that pinning is very strong. Finally, we use bulk B phase values for parameters $\rho_\mathrm{s}$, $\Delta$ and $\xi$ to evaluate the expressions (\ref{pinning_estimate},\ref{Magnus_force}) as a function of temperature for simplicity. 

The mutual friction parameters $\alpha$ and $\alpha'$ are likely also modified with respect to the A and B bulk phases, say, because of the aerogel strands penetrating the vortex cores and modifying the core-bound states' spectra \cite{kopnin1991mutual}. However, we expect the general temperature dependence of the coefficients to remain similar to that in the bulk phases. Between the bulk phases, the differences at temperatures down to $\sim 0.5 \,T_\mathrm{c}$ are not very large \cite{bevan1997vortex}, and $\alpha\to\infty$ and $\alpha'\to1$ as $T\to T_\mathrm{c}$ \cite{bevan1997vortex}. We use $\alpha = 20 , \alpha' = 1$ for a simulation of qualitative behaviour very close to $T_\mathrm{c}$ and $\alpha = 0.1 , \alpha' = 0.2$ for a simulation at $T\sim 0.5 \, T_\mathrm{c}$ where most of the experiments were carried out. We simulate both SQVs and HQVs with the same model, just changing the circulation quantum accordingly. The extra pinning of HQVs due to the soliton structure in the spin part of the order parameter connecting adjacent HQVs in pairs \cite{HQVs_prl,Mineev2014,VolovikMineev1976} is neglected. 

%%%%%%%%%%%%%%%%%%%%%%%%%%%%%%%%%%%%%%%%%%%%%%%%%%%%%%%%%%%%%%%%%%%%%%%%%%%%%%

\begin{figure}[t]
\includegraphics[width=0.95\linewidth]{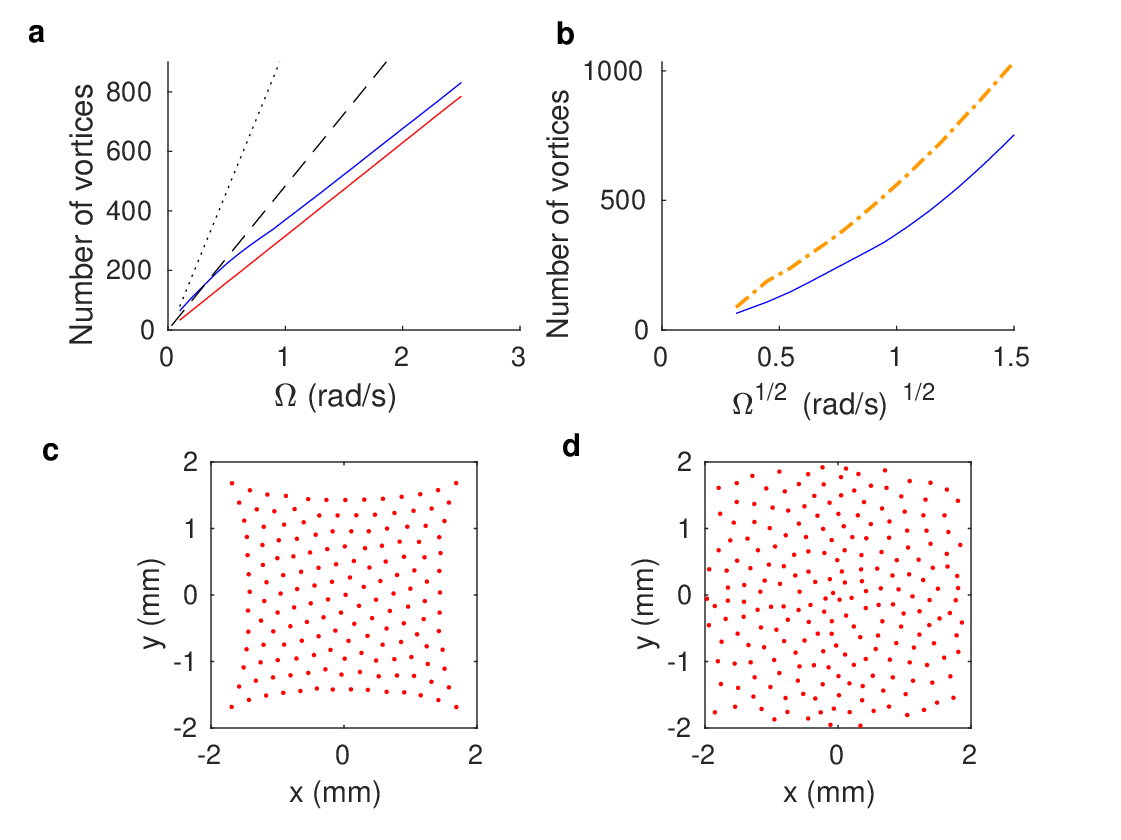}%
\caption{Simulated dynamic steady state number of single-quantum vortices in the core-like aerogel as a function of rotational velocity, $\Omega$: ({\bf a}) The simulation starts from a uniform array of vortices (black dotted line) with a density at least twice as high as in the final stable configuration. Vortices are allowed to move until a stable vortex number is established. The steady-state vortex number in the box-shaped container very close to $T_\mathrm{c}$ ($\alpha=20$, $\alpha'=1$, red line) with no pinning is proportional to the rotation velocity. It becomes nonlinear in $\Omega$ if weak pinning is turned on ($v_\mathrm{c,p}=0.1$\,mm/s, blue line). This line is obtained by averaging over 10 simulations with 100\,$\mu$m positional noise in the initial condition. The black dashed line shows the equilibrium number of vortices in a cylinder with the same surface area.
% The black dotted line is the number of vortices in the initial, uniform state.
({\bf b}) The steady-state vortex number with weak pinning is roughly linear as a function of $\sqrt{\Omega}$ (blue line) up to $\Omega=1$\,rad/s. This result depends on the mutual friction parameters. Decreasing them to $\alpha=2$, $\alpha'=0.8$ and keeping $v_\mathrm{c,p}=0.1$\,mm/s (orange dashed line) extends the linearity in $\sqrt{\Omega}$ up to the highest rotation velocity measured in the experiments. ({\bf c}) The final vortex configuration at a temperature very close to $T_\mathrm{c}$ (red line in panel (a) at $\Omega=0.5$\,rad/s) shows voids where the superflow enhancement by the box geometry is the largest. ({\bf d}) These voids are filled in the pinned state corresponding to the blue line ($\Omega=0.5$\,rad/s), which results in the larger total vortex number. \label{FIG:SQV_number}}
\end{figure}

%%%%%%%%%%%%%%%%%%%%%%%%%%%%%%%%%%%%%%%%%%%%%%%%%%%%%%%%%%%%%%%%%%%%%%%%%%%%%%

Based on these assumptions, we can now understand the experimental features outlined above. We first focus on the $\sqrt{\Omega}$ scaling in the relaxation rate for SQVs. In Fig. \ref{FIG:SQV_number} we simulate the system, starting with a density of vortices at least twice as high as in the final stable configuration, and allow the configuration to evolve until the vortex number is no longer decreasing. We note that this approach gives a finite final vortex number at zero rotation, and the effective depinning velocity where the configuration freezes during cool-down decreases with the rotation velocity. As a compromise, the simulation outcome is not shown below $\Omega=0.1$\,rad/s, which is approaching the rotation velocity where no isolated vortex can be depinned in our container with $v_\mathrm{c,p}=0.1$\,mm/s. 

In the case without pinning we observe that the high temperature steady-state vortex number is proportional to the rotation velocity, just like in a cylindrical container, albeit somewhat lower owing to the voids near the flow maxima at the box wall centres (see panel (c) in Fig.~\ref{FIG:SQV_number}). This outcome does not depend on the mutual friction coefficients as long as $\alpha>\alpha'$.  Remarkably, if we add weak pinning ($v_\mathrm{c,p}=0.1$\,mm/s) to mimic the onset of pinning when the system has just passed through $T_\mathrm{c}$ in rotation, the steady-state vortex number is linear in $\sqrt{\Omega}$ up to $\Omega=1$\,rad/s, owing to the added vortices the system with weak pinning is able to hold in the voids (see panel (d) in Fig. \ref{FIG:SQV_number}). This outcome depends on the mutual friction parameters: decreasing the coefficients to $\alpha=2$, $\alpha'=0.8$ extends the linearity in $\sqrt{\Omega}$ up to $\Omega=2.5$\,rad/s, which is the maximum rotation velocity applied in the experiments \cite{HQVs_prl}. When the temperature decreases further, $v_\mathrm{c,p}$ quickly increases and further changes to the vortex configuration are thus prohibited. These aspects provide a plausible explanation for the experimental observation that $1/\tau(\Omega) - 1/\tau(0) \propto \sqrt{\Omega}$. That is, the relaxation rate is proportional to the total vortex number (line length) as expected, but the vortex number coincidentally scales as $\sqrt{\Omega}$. This result removes the ambiguity of interpreting spin-up and spin-down experiments in core-like aerogel where SQV dynamics are followed using this relaxation rate. Our conclusions further substantiate the validity of the pinned vortex model used here and in the previous section. 

If the dynamic steady state for SQVs with pinning follows $N \propto \sqrt{\Omega}$, this raises the question why the number of HQVs produced is $N \propto \Omega$ as reported in \cite{HQVs_prl}. Repeating the simulation in Fig.~\ref{FIG:SQV_number} for HQVs, by halving the circulation quantum, yields a similar result with the vortex numbers simply doubled. One plausible reason is that the critical velocity for adding SQVs to the system is lower than that for adding HQVs, which has been experimentally confirmed around $T=0.4T_\mathrm{c}$ \cite{PhysRevResearch.2.033013}. The $\sqrt{\Omega}$-dependence for SQVs simulated above arises because the total steady-state vortex number with pinning turned on is larger than that with pinning off. That is, vortices need to be added to the system once pinning becomes relevant as the temperature decreases. If the vortices added at temperatures where pinning applies are SQVs only, the HQV number is determined very close to $T_\mathrm{c}$ where pinning is still irrelevant, before the vortex configuration freezes. Just before freezing, the steady-state vortex number is $\propto \Omega$ as shown by the red line in Fig.~\ref{FIG:SQV_number}a calculated in the absence of pinning. This qualitative outcome does not change if we halve the circulation quantum. With these assumptions, the HQV number would thus scale $\propto \Omega$.

Let us now focus on the production of vortices by the KZ mechanism, pinned in place by the aerogel. The KZ mechanism produces all vortex types independently with the characteristic inter-vortex distance of $l_\mathrm{KZ}= a \xi_0 (\tau_\mathrm{Q}/\tau_0)^{1/4}$, where $\tau_0= \xi_0/v_\mathrm{F}$, $a$ is a dimensionless parameter of the order of unity, $v_\mathrm{F}$ is the Fermi velocity, and $\tau_Q$ is the cool-down rate in units of $T_\mathrm{c}$. The typical KZ vortex line length in \cite{HQVs_prl, PhysRevLett.127.115702} corresponds to roughly the dynamic steady-state vortex configuration at $\Omega=1$\,rad/s, but with random orientations \cite{HQVs_prl}. Even for the slowest cool-down through the superfluid transition reported in \cite{HQVs_prl, PhysRevLett.127.115702}, the system reaches $T=(1-10^{-5}) T_\mathrm{c} $ in a few seconds. This can be compared with the estimate for $v_\mathrm{c,p}$ obtained from Eqs.~(\ref{pinning_estimate},\ref{Magnus_force}) using B phase parameters. The applicable filling factor for vortices along the pinning sites is $\sigma=1$, but the KZ vortices are randomly oriented. As a pessimistic estimate, we thus use $\sigma\approx1/5$, corresponding to the lateral spacing of the rods. The obtained depinning velocity goes to zero at $T_\mathrm{c}$, while it reaches $v_\mathrm{c,p}\sim 4$\,mm/s at $T=0.999 T_\mathrm{c}$ and is still $v_\mathrm{c,p}\sim 0.4$\,mm/s at $T=(1-10^{-5}) T_\mathrm{c}$. 

\begin{figure}[t]
\includegraphics[width=1\linewidth]{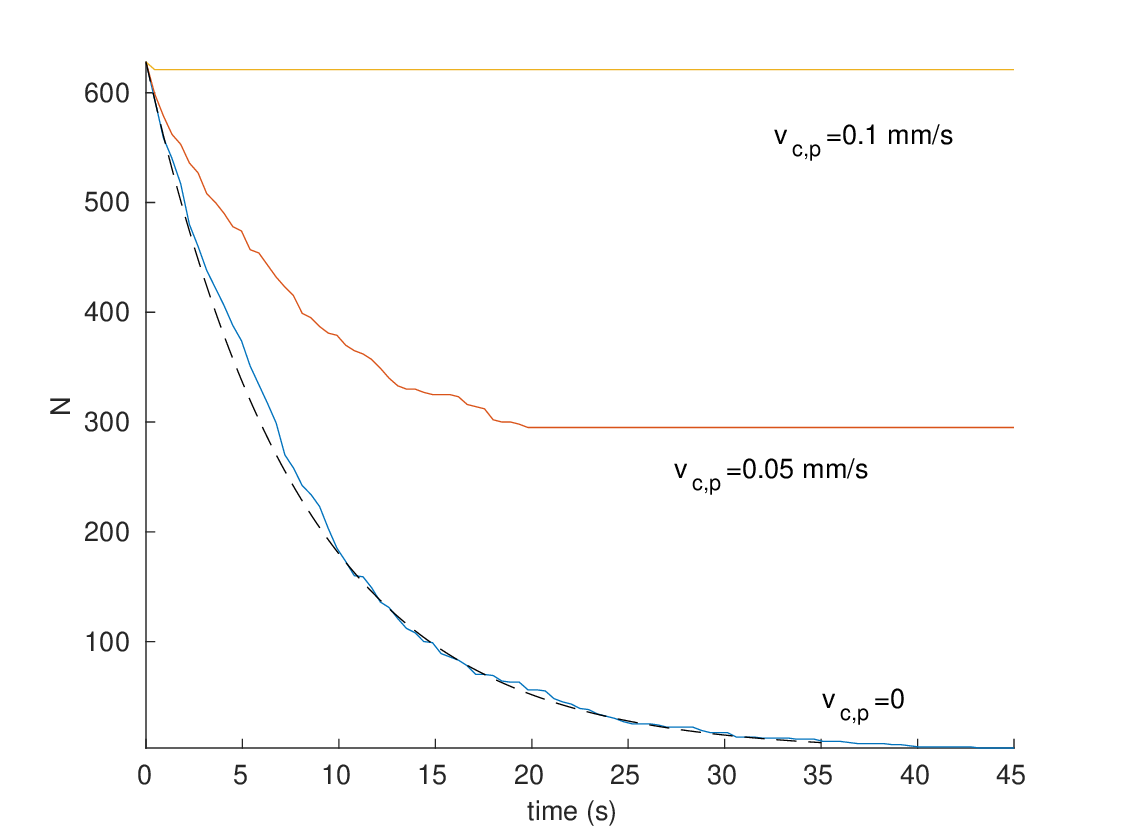}%
\caption{Simulated decay of HQVs created by the Kibble-Zurek mechanism in core-like aerogel: The simulation starts from a vortex array corresponding to $\Omega=1$\,rad/s where the vortex polarities have been randomised to mimic a typical vortex density created by the KZ mechanism. In the absence of pinning ($v_\mathrm{c,p}=0$, blue line), the vortices move out of the container and annihilate one another exponentially with a time constant of $8$\,s (black dashed line). When pinning is turned on, the decay time constant does not change but some vortices remain in place in the final steady state. If the depinning velocity is $v_\mathrm{c,p} \ge 0.1$\,mm/s, no vortices are annihilated from the initial state. Here $\alpha'=1$ and $\alpha =20$, corresponding to a temperature very close to the superfluid transition temperature. Similar results apply to SQVs with only the vortex numbers halved due to their circulation quantum that is twice as large.  \label{FIG:HQV_decay}}
\end{figure}

The experimental observation that KZ vortices freeze in due to pinning can be tested against a point vortex simulation. The simulation starts from a random array of positive and negative vortices corresponding to the steady-state number of vortices at $\Omega=1$\,rad/s (Fig. \ref{FIG:HQV_decay}). We find that the removal of vortices by mutual annihilation and by motion out of the container takes place exponentially with a time constant of $8$\,s, and no vortices are removed if $v_\mathrm{c,p} \geq 0.1$\,mm/s. The time constant is coarsely doubled if only annihilation is allowed, which we probe by moving the outer boundary to infinity after the initial configuration is created. Here we used $\alpha=20$ and $\alpha'=1$, although increasing $\alpha$ makes the dynamics slower (increasing $\alpha'$ makes it faster, but $\alpha'$ already has its maximum value). That is, this simulation provides a pessimistic estimate of the near-$T_\mathrm{c}$ pinning capacity. These results thus demonstrate how the vortices produced by the KZ mechanism can freeze in without decaying. The quantitative details should be taken as rough estimates only, because the pinning model used may not be reliable close to $T_\mathrm{c}$ where the vortex cores envelop several pinning sites, and because we used bulk B phase parameters to evaluate the temperature dependence of $v_\mathrm{c,p}$. However, drawing the above observations together, we conclude the pinning force estimate Eq.~(\ref{pinning_estimate2}) is consistent with the experimental evidence from experiments close to the superfluid transition temperature in the core-like aerogel.

%%%%%%%%%%%%%%%%%%%%%%%%%%%%%%%%%%%%%%%%%%%%%%%%%%%%%%%%%%%%%%%%%%%%%%%%%%%%%%

%%%%%%%%%%%%%%%%%%%%%%%%%%%%%%%%%%%%%%%%%%%%%%%%%%%%%%%%%%%%%%%%%%%%%%%%%%%%%%
%%%%%%%%%%%%%%%%%%%%%%%%%%%%%%%%%%%%%%%%%%%%%%%%%%%%%%%%%%%%%%%%%%%%%%%%%%%%%%

\begin{figure}[t]
\includegraphics[width=1\linewidth]{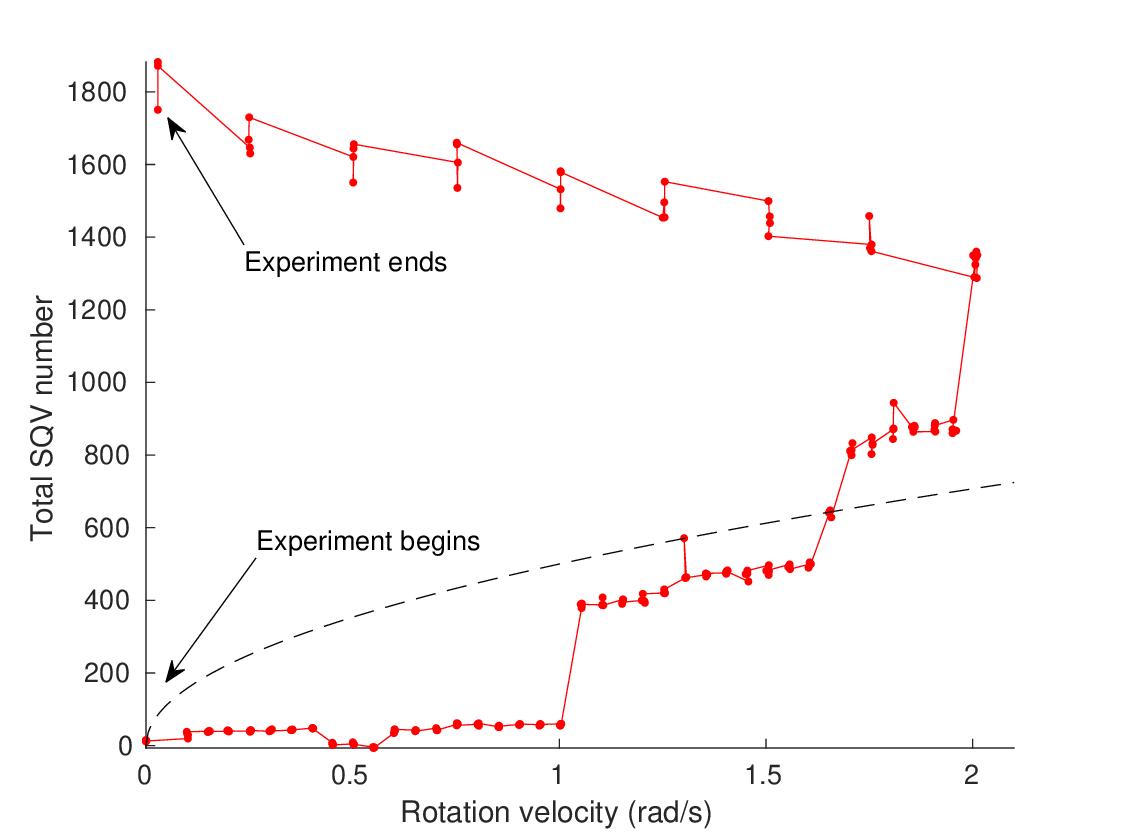}%
\caption{Experimental dynamics of SQVs in the core-like aerogel: The red points connected by a line show a canonical experiment. The experiment, prepared to exclude KZ vortices, begins by cooling down in the absence of rotation.  The rotation velocity is then increased in small steps, and changes in the SQV line length are monitored by measuring changes in the relaxation rate of a magnon condensate in the same volume. The relaxation rate $1/\tau$ is converted to the line length (vortex number) assuming $1/\tau(\Omega) - 1/\tau(0) \propto N \propto \sqrt{\Omega}$. In the conversion the initial relaxation rate at $\Omega=0$ is subtracted for clarity. The scale of the y axis is obtained from the slope of the orange line in Fig.~\ref{FIG:SQV_number}b (combined with the blue line in Fig.~3b in \cite{PhysRevResearch.2.033013} to convert relaxation rate to rotation velocity), but the argument remains the same if a different slope is used. The first significant increase in the SQV number takes place at $\Omega\approx 1$\,rad/s. The black dashed line is the measured vortex number in a dynamic steady-state array of SQVs ($\propto \sqrt{\Omega}$ as discussed in the text), showing that the total vortex line length quickly exceeds the steady-state value. Once $\Omega \approx 2$\,rad/s is reached, the rotation velocity is gradually reduced back to zero. The total vortex line length keeps increasing, demonstrating that vortex motion out of the container is not how the system responds to changes in $\Omega$. No HQVs are produced in this experiment, as confirmed by independent measurements of the continuous-wave NMR spectrum. The experimental data is taken from Fig. 5 in \cite{PhysRevResearch.2.033013}. \label{FIG:SQV_dynamics}}
\end{figure}

%%%%%%%%%%%%%%%%%%%%%%%%%%%%%%%%%%%%%%%%%%%%%%%%%%%%%%%%%%%%%%%%%%%%%%%%%%%%%%

\subsection*{Interpreting spin-up and -down experiments}

We next turn our attention to the process of vortex generation in the core-like aerogel with rotation initiated after cooling well into the superfluid state, recreating the dynamics also taking place during neutron star spin-down. In the experiment reported in \cite{PhysRevResearch.2.033013}, the rotation velocity is changed in small steps, and variations in the SQV line length are monitored by measuring changes in the relaxation rate of a magnon condensate penetrated by the vortices. We convert the measured relaxation rate $1/\tau$ to the SQV line length (vortex number) using the dependence $1/\tau(\Omega) - 1/\tau(0) \propto N \propto \sqrt{\Omega}$ substantiated above. We argue that the observed spin-up and spin-down dynamics are qualitatively different from the process of vortex depinning and motion that takes place in the crust-like aerogel outlined above. There are two key aspects in the data that differentiate the core-like aerogel from the crust-like one.

First, as the rotation velocity spins up from zero, the first SQVs enter the system at $\Omega\approx1$\,rad/s (Fig.~\ref{FIG:SQV_dynamics}) in a sudden avalanche. HQVs are not produced as confirmed by the absence of the satellite line. This critical rotation velocity corresponds to a superflow velocity of 2\,mm/s at the middle of the box side walls, where the flow is the fastest. This $v_\mathrm{c}$ is smaller than that observed in the crust-like aerogel, while the critical unpinning flow velocity calculated in the same manner as outlined for Fig.~\ref{FIG:pinning_forces} but using core-aerogel parameters ($d=10$\,nm, $T=0.4 T_\mathrm{c}$, $p=7$\,bar, $\sigma\approx 1$) yields $v_\mathrm{c,p}\sim 10^2$\,mm/s. This value is much larger than in the crust-like case, mostly owing to the pinning sites being continuous and aligned with the vortices. Thus, the process that increases the vortex number in the core-like aerogel experiment is not large-scale depinning of vortices flowing in from the perimeter as we observed in the crust-like system. 

Second, as the changes in the relaxation rate are proportional to the number of SQVs, we can compare the steady-state vortex number obtained by cooling down slowly in rotation, taken from Fig.~3 in \cite{PhysRevResearch.2.033013}, with the observed increase in a spin-up experiment from Fig.~5 in the same paper (Fig.~\ref{FIG:SQV_dynamics}). The measured relaxation rate is first converted to the corresponding equilibrium rotation velocity using the blue line in Fig.~3b in \cite{PhysRevResearch.2.033013}. The equilibrium vortex number at a given rotation velocity is then set by taking the slope of the orange line in panel b in Fig.~\ref{FIG:SQV_number}. However, we emphasise that the absolute scale of the y axis in Fig.~\ref{FIG:SQV_dynamics} is unimportant for the comparison of the dynamic data with the equilibrium vortex array. In the spin-up experiment the observed vortex number exceeds the dynamic steady-state vortex number by a factor of two at $\Omega=2$\,rad/s. Moreover, the vortex number does not decrease as the rotation velocity is subsequently decreased back to zero, but rather increases further. This indicates that the system reacts to changes in the rotation velocity not by depinning vortices and moving them in or out of the container but instead by creating new vortices. During spin-down, the added vortices carry the opposite polarity to those added in the spin-up. Some of the pinned vortices get annihilated in this process, as indicated by the total number not being doubled in the spin-down.

A natural interpretation of these observations is illustrated by the toy model simulations in the original article \cite{PhysRevResearch.2.033013} and qualitatively supported by the phenomenology of vortex avalanches observed in superconducting films \cite{altshuler2004colloquium}. That is, the observed bursts of vortices enter as disordered branches, spreading inwards from the points where the superflow first reaches the critical value $v_\mathrm{c}\approx2$\,mm/s at the container side wall centres. The microscopic process creating these branches remains unclear. It may be vortex ring nucleation and expansion in the bulk superfluid, initiated when $v_\mathrm{c,n}$ is exceeded \cite{ruutu1997intrinsic}, or straight vortex nucleation at the pinning sites, made possible by the flow enhancement around the pinning sites. Or it may be vortices entering from the top and bottom of the sample, presumably via the Kelvin-Helmholtz instability \cite{eltsov2024kelvin,eltsov2019kelvin} at the phase boundary between the polar phase in the aerogel and the B phase between the container wall and the aerogel edge. The vortices in this last scenario would subsequently expand vertically as permitted by the potentially weaker pinning for vortex motion along the cylindrical pinning sites. In a perfectly uniform aerogel the vertical pinning force would be zero. 

In principle, vortices may also enter through the sides of the container in bursts sustained by the local changes in the flow environment once the first vortices enter the system. Running our simulation model using the high-temperature pinning estimate $v_\mathrm{c,p}=20$\,mm/s obtained above shows that during the spin-up vortices cannot enter into the aerogel even if the total vortex number is allowed to become 10000. This is because the vortex spacing needs to be of the order of $100$\,nm before adjacent vortices can unpin one another, meaning that these vortices cannot be pushed into the aerogel even if they all get inserted from a single point, say, at the side wall centres. Therefore, this mechanism would only accumulate vortices at the very edges of the aerogel. In such a highly localised vortex configuration, the magnon condensate relaxation rate near the boundary would be much faster than elsewhere. Hence, the longest-lived signal (which is used to extract the vortex configuration \cite{PhysRevResearch.2.033013,autti2018bose}) would probe the vortex configuration away from the edge of the container, thus not leading to the reported decrease in the life time. Therefore, we assess this sideways motion scenario to be unlikely. On the other hand, the processes involving vortex nucleation or vortex motion in the third dimension cannot be modelled within the two-dimensional point vortex model, and thus resolving the microscopic mechanism remains an interesting task for the future.
 
Putting these arguments together, we conclude that in the core-like aerogel $v_\mathrm{c,n}<v_\mathrm{c,p}$. The observed critical velocity is related to the intrinsic nucleation of vortices in the bulk of the system, or to vortices entering through the top and the bottom of the aerogel and extending along the vertical dimension (which has the same qualitative effect), when the local superflow velocity exceeds $v_\mathrm{c}\approx 2$\,mm/s. The vortex depinning velocity is likely an order of magnitude larger than this, but it cannot be extracted from existing data \cite{PhysRevResearch.2.033013,PhysRevLett.127.115702}. We note that whether or not vortices can enter through the sides of the aerogel is irrelevant as long as vortex depinning and motion are not allowed.

%%%%%%%%%%%%%%%%%%%%%%%%%%%%%%%%%%%%%%%%%%%%%%%%%%%%%%%%%%%%%%%%%%%%%%%%%%%%%%

\section*{Implications for Neutron Stars}

From the above considerations, we conclude that the depinning superflow velocity can be reliably estimated from the pinning force of Eq.~(\ref{pinning_estimate}) and the Magnus force (\ref{Magnus_force}) in two different superfluid $^3$He-aerogel environments and at a range of temperatures, including those not accessible in steady-state experiments. Our work outlines two phenomenologically different pinned vortex scenarios, one where superflow is able to eventually move vortices and another where intrinsic vortex nucleation facilitates responses to changes in the flow environment. The division of the aerogel experiments into crust-like and core-like has so far been based on geometric similarities with the underlying structures in neutron star superfluid phases. Let us now extend the analogy by applying the obtained pinning force and vortex dynamics phenomenology to the neutron star superfluids. Relevant physical parameters to translate between the systems are listed in Table \ref{table:comparison}.

The most important observation from the crust and core-like aerogel experiments is that the vortex pinning force provided by these structures is significantly larger than that calculated from the straightforward condensation energy gain from vortex core displacement. That is, Eq.~(\ref{pinning_estimate2}), argued elsewhere \cite{Mesoscopic16} to be an overestimate of the available pinning force in the neutron star crust by several orders of magnitude based on averaging the interaction between straight vortices and different lattice orientations, is actually an underestimate by two orders of magnitude in superfluid $^3$He. The reason why the directional averaging of the pinning force becomes irrelevant in aerogel experiments is that the pinning force (energy), Eq.~(\ref{pinning_estimate}), is very large, thus reducing the orientation-averaging suppression factor $2 \mathcal{T} R_\mathrm{ws}/|E_\mathrm{site}|$ (see expression (15) in \cite{Mesoscopic16}) from $\sim10^3$ to $\sim 1$. Here $R_\mathrm{ws}$ is the typical lateral distance between pinning points (in the crust-like aerogel), $\mathcal{T}$ is the vortex tension and $E_\mathrm{site}$ is the pinning energy associated with Eq.~(\ref{pinning_estimate}) per pinning site. That is, vortices are not sufficiently rigid compared to the strong pinning force presented here and will bend and adapt to the pinning landscape. Hence, the spatial averaging does not wash out the local pinning effects, and the pinning force of Eq.~(\ref{pinning_estimate}) applies in full. Let us now apply this vortex pinning picture in the neutron star context.

For the stellar core, Eq.~(\ref{pinning_estimate}) yields $v_\mathrm{c,p}\sim 5 \times10^6$\,m/s, which corresponds to a large lag in rotational speeds between the superfluid and normal component of 550\,rad/s (at the outer edge of a neutron star core of radius 9\,km) and therefore requires rapid neutron star rotation. For the pinning force estimate, we set $\sigma=1$ corresponding to flux tubes aligned with the rotation axis, and used a fiducial density $\rho=10^{18}$\,\SI{}{\kilogram \,\metre^{-3}} (see Table \ref{table:comparison}). Picking instead a worst-case filling factor of $\sigma=0.1$ corresponding to very tilted flux tubes of 100\,fm diameter and 1\,pm lateral spacing yields $v_\mathrm{c,p}\sim 5 \times10^5$\,m/s (equivalent to a lag of 55 \,rad/s). Our small-scale simulations in Figs.~\ref{FIG:cylinder_pinning_simulation}--\ref{FIG:cylinder_pinning_simulation2} show that vortex unpinning does not happen before the superflow felt by the vortex reaches $\sim v_\mathrm{c,p}$. During spin down, the required flow is accumulated as the rotation velocity of the non-superfluid components including the pinned vortices decelerates while the superflow (created by the pinned vortices) is not changing. Provided these simulation results can be extrapolated to neutron star scales and given the broad range of neutron star rotation frequencies ($\sim 0.5 - 500$ \,rad/s), even the latter pessimistic estimate of the core pinning  allows a wide window for a potential vortex nucleation mechanism to act before unpinning could take place. That is, in a bulk superfluid nucleation of vortices via hydrodynamic instability, where the energy barrier for vortex production is reduced to zero, is expected coarsely at $v_\mathrm{c,n}\sim \kappa /(2 \pi \xi) \sim 10^6$\,m/s  (see Fig.~26.7 in \cite{VolovikBook}). Immersing coherence-length-scale solid structures into the bulk superfluid decreases $v_\mathrm{c,n}$ due to the resulting flow enhancement around the immersed structures. We also note that, for a neutron star that rotates slower than the threshold implied by $v_\mathrm{c,p}$ when passing through the superfluid transition, vortex depinning is unavailable in a spin down all the way to zero rotation, thus inevitably calling for a mechanism other than vortex depinning to explain the observed rotation glitches \cite{Basu2022}.

For the stellar crust, the estimated unpinning velocity is larger than $v_\mathrm{c,p}\sim 10^5$\,m/s. Here, we have used $\sigma=0.1$ as a lower-bound estimate, corresponding to 100\,fm pinning site spacing and 10\,fm pinning site radius, as well as $\rho=10^{17}$\,\SI{}{\kilogram \,\metre^{-3}} and $\xi=100$\,fm corresponding to the base of the crust \cite{graber2018glitch}. This unpinning velocity corresponds to coarsely 10\,rad/s solid-body rotation at a 10\,km radius. With this $v_\mathrm{c,p}$, the inner crustal layers in many neutron stars are probably able to react to a spin down by moving vortices outwards. However, choosing $\rho=10^{15}$\,\SI{}{\kilogram \,\metre^{-3}} and $\xi=10$\,fm corresponding to the outer edge of the superfluid region in the inner crust \cite{graber2018glitch} yields $v_\mathrm{c,p}\sim 10^7$\,m/s ($\sigma=0.1$), or 1000\,rad/s, showing that the vortex motion may get impeded before the vortices are able to move out of the crustal superfluid for stars that are born not too rapidly rotating. In this scenario, the vortex configuration would freeze during spin down with all the vortices packed in the outer layers of the inner crust. Thus, bulk vortex nucleation may be relevant also in the neutron star crust, which is characterised by roughly the same $v_\mathrm{c,n}$ as the stellar core.

%The motion of negative polarity vortices along the axis of rotation (or pinning sites) from the outer edge, corresponding to $v_\mathrm{c,e}$, seems unimportant in a neutron star simply due to the large distance that the vortices would need to travel to expand along the length of the star, but a more detailed discussion of this possibility is left for the future. 

We note that under the strong pinning scenario outlined above a range of bulk superfluid phenomena become irrelevant for spin-down physics in a neutron star once the temperature is well below the superfluid transition temperature. Most of the time the vortices are immobile. When the superflow changes, this is either because pinned vortices are gradually pushed from one microscopic pinning position to the next (as in the crust-like aerogel) or because new vortices are nucleated in the bulk of the system and immediately pinned in place (as in the core-like aerogel). There is no large-scale vortex motion, and therefore no turbulence of the vortex array either. We thus speculate that the pinned vortex simulation techniques developed in the crust-like section above can be directly applied to neutron star superfluids as well, and owing to the strong pinning, these simulations can potentially be scaled up well beyond existing simulations \cite{howitt2020simulating} by, say, only updating the positions and flow contributions of those vortices that are outside a threshold radius for vortex motion (see Fig.~\ref{FIG:cylinder_pinning_simulation}). Phenomenological simulations of vortex nucleation avalanches in a neutron star (and the core-like aerogel) will require developing new simulation techniques, but the immobility of the vortices once created opens an avenue for supersizing these simulations as well. We also highlight the strong pinning available essentially continuously on the scale of any conceivable intervortex distance throughout the superfluid phases, significantly simplifying applicable numerical simulations, as demonstrated by the superfluid $^3$He simulations developed in this Article.

Finally, when the neutron star superfluid phases are created as the star cools down, the KZ mechanism produces all possible types of topological defects. These topological defects may get pinned in place before they can decay, as demonstrated by the core-like aerogel experiments. The aerogel experiments also show that KZ vortices may be created additively, on top of those imposed by rotation \cite{PhysRevLett.127.115702}. In such circumstances, the KZ vortices create a random landscape of superflow and pinned vortices. This randomness may then seed different spatial vortex dynamics patterns as the rotation velocity decreases, potentially giving rise to the spectrum of glitch behaviour observed in the known pulsars.

 \begin{turnpage}
\begin{table*}[b!]
  \centering
  \caption{Characteristics of superfluid $^3$He in terrestrial experiments vs neutron star (NS) superfluidity. We distinguish crust-like and core-like aerogels and provide relevant parameters for the NS crust and core separately. 
  % In superfluid $^3$He experiments, the characteristic parameters can be adjusted using pressure 0$\le p \le$ 34\,bar, temperature, Nafen aerogel packing density, etc. 
  The atomic mass unit is denoted as $m_0$, $r$ is the transverse distance from a vortex, and $\rho_s$ denotes the superfluid mass density.
  % the relative density of the superfluid component is $\rho_s$. 
  For the core-like aerogel experiments, $^3$He values are given for the pressure and temperature at which the largest body of applicable experiments have been carried out, which is $P=7.1$\,bar and $T/T_c=0.4$ for the core-like aerogel and $P=30$\,bar and $T/T_c\sim0.5$ for the crust-like aerogel. For the NS parameters, we focus on those stars that are older than $\sim 10^3\,$yrs and sufficiently cold to host superfluid interiors. Unless stated otherwise, we focus on the perspective of the neutron superfluid.}
  \vspace{0.3cm}
  \begin{tabular}{c||c|c|c|c}
    & $^3$He in crust-like aerogel & $^3$He in core-like aerogel & NS crust & NS core   \\
    \hline 
    \hline
    pairing state           & p-wave (B phase) & p-wave (polar phase)
        & s-wave (neutrons) & s-wave (protons), p-wave (neutrons) \\      
    % \hline 
    % pressure & & $P=7$\,bar & & \\    
    \hline 
    transition temperature, $T_c$ &2.4\,mK& $\sim$ 1\,mK & $\sim 10^{10}\,$K (neutrons) & $\sim 10^{10}\,$K (protons), $\sim 10^{9}\,$K (neutrons) \\
    \hline 
    ambient temperature, $T$ & $0.4 \ge T/T_c \ge 1$ & $0.1 \ge T/T_c \ge 1$ & $T \ll T_c$ & $T \ll T_c$ \\    
    \hline 
    coherence length, $\xi_0$ & 20\,nm & $40$\,nm & $\sim 10 - 100 \,$fm & $\sim 100 \,$fm \\    
    % \hline 
    % superfluid gap & & $\Delta \approx 1.76 k_B T_c$ & & \\    
    % \hline
    % density of states & & $N_0 = 6.8\cdot  10^{50}$ \SI{}{J^{-1}m^{-3}} & & \\   
    \hline 
    quantum of circulation, $\kappa$ & $\approx h/(6 m_0)$ & $ \approx h/(6 m_0)$ 
        & $\approx h/(2 m_0)$ & $\approx h/(2 m_0)$ \\    
    \hline 
    mass density, $\rho$ & \SI{115}{\kilo\gram\, \meter^{-3}} & \SI{95}{\kilo\gram\, \meter^{-3}} & $10^{15 - 17}$ \SI{}{\kilo\gram\, \meter^{-3}} & $10^{17-18}$ \SI{}{\kilo\gram\, \meter^{-3}} 
    \\    
    % \hline 
    % Fermi momentum & & $p_F = 8.7\cdot 10^{-25}$ \SI{}{\kilo\gram\, \meter\, \second^{-1}} & & \\    
    % \hline 
    % Fermi velocity & & $v_F = 50$ \SI{}{\meter\, \second^{-1}} & & \\
    % \hline 
    % Landau velocity & & $v_L=0$ & $v_L\approx 10^7$ cm/s \cite{Allard23} & same \\ 
    \hline  
    relevant scale & $\sim 5\,$mm & $\sim 5\,$mm 
        %\cite{HQVs_prl}
        & $\sim 1\,$km  & $\sim 10\,$km \\
    % \hline  rotation velocity & $|\Omega|= 0 - 2.5$ rad/s  &$\Omega=10-100$ rad/s \footnote{for typical glitching pulsars, faster pulsars generally are recycled, spin down slower, and do not exhibit many glitches}& generally the same as the crust \footnote{one does not generally assume a difference in rotation between crust and core, but rather between two compenetrating fluids, one normal one superfluid, which may be in the crust or core, depending on the nature of the pinning.} \\
    % \hline  acceleration & $|\dot{\Omega}|\le$ \SI{30}{\milli\radian\, \second^{-2}}  & \SI{0.01}{\milli\radian\, \second^{-2}}$\lesssim \dot{\Omega}\lesssim$ \SI{100}{\radian\, \second^{-2}}& \cite{VelaRise, WM12} same \footnote{our simulations give $\dot{\Omega}\approx 10^{-4}-10^{-3}$ rad/s$^2$ \, \cite{Haskell2012}} \\
    % 
    \hline 
    inter-vortex distance  & $\gtrsim$50\,$\mu$m & $\sim$100\,$\mu$m %\cite{PhysRevLett.127.115702} 
        & $\sim$100\,$\mu$m (for $P\sim 1\,$s) %\cite{haskell2015models}
        & $\sim$100\,$\mu$m (for $P\sim 1\,$s)\\
    \hline 
    vortex pinning cause & silica aerogel & Nafen aerogel & crustal lattice  & superconducting fluxtubes  \\
    \hline
    pinning site geometry & spherical (strands) & cylindrical &  spherical & cylindrical \\
    \hline 
    pinning site diameter, $d$ & $\sim 5\,$nm & $\sim 10$\,nm\, %\cite{dmitriev2020superfluid}
        & $\sim 10\,$fm
        % \cite{Mesoscopic16} 
        & $\sim 100\,$fm 
        %\cite{Link03}
        \\
    \hline pinning site spacing & $\sim 100\,$nm & $\sim 50$ \,nm\, 
        %\cite{dmitriev2020superfluid}
        & $\sim 10-100\, $fm 
        % \cite{Mesoscopic16}
        & $\sim 1\,$pm (for $B\sim 10^{12}\,$G)
        % \cite{GAS} 
        \\
    % \hline 
    % pinning force per unit length & & $F_p \sim \rho_s N_0 \pi \Delta^2 d $  &  &   \\
    % \hline 
    % Magnus force per unit length & & $F_m = \kappa^2 \rho \rho_s \pi /r  $  &  &   \\
    % \hline 
    % releasing transverse pinning & & $v_{c1}\gtrsim$ \SI{20}{\milli\meter\, \second^{-1}} \footnote{to be explained in this paper} & &  \\ 
    % \hline 
    % eff. pinning site spacing & & $\sim$ \SI{40}{\micro\meter} \footnote{to be explained in this paper} & & \\
    % \hline 
    % vortex formation from bulk\cite{VolovikBook} & & $v_{c2}\gtrsim \SI{50}{\milli\meter\, \second^{-1}}$ & &  \\     
    % \hline 
    % loop emission & & $v_{c3}\gtrsim \SI{50}{\milli\meter\, \second^{-1}}$ & &  \\  
    % \hline 
    % turbulence  & &  $v_{c4}\gtrsim \SI{40}{\milli\meter\, \second^{-1}}$ & &  \\      
    % \hline 
    % Kelvin-Helmholtz instability & & $v_{c,KH}\sim \SI{2}{\milli\meter\, \second^{-1}}$  &  &  \\ 
    % \hline 
    % Kibble-Zurek vortex distance & & $l_{KZ} = \xi_0 (\tau_Q/\tau_0)^{1/4} \sim$\,\SI{4}{\micro\meter}  &  &   \\ 
    % \hline 
    % cool-down rate at $T_c$ & & $\tau_Q^{-1} \sim $ $ \frac{T/T_c}{10^3\, \mathrm{s}}$ - $ \frac{T/T_c}{10^5\, \mathrm{s}}$ &  &  \\ 
    % \hline 
    % order par. rel. rate & & $\tau_0 = \xi_0 v_F$  &  &  \\ 
    % \hline 
    %  typical KZ equiv. & $\Omega$ & 0.5 - 1\,rad/s &  &  \\ 
  \end{tabular}\label{table:comparison}
\end{table*}
 \end{turnpage}

%%%%%%%%%%%%%%%%%%%%%%%%%%%%%%%%%%%%%%%%%%%%%%%%%%%%%%%%%%%%%%%%%%%%%%%%%%%%%%
%%%%%%%%%%%%%%%%%%%%%%%%%%%%%%%%%%%%%%%%%%%%%%%%%%%%%%%%%%%%%%%%%%%%%%%%%%%%%%

\begin{acknowledgments}
We thank Risto Hänninen for the original cylinder-geometry 2D point vortex simulation codes that were modified to model vortex pinning and, also, the box geometry. We also thank Vladimir Eltsov, Jere Mäkinen and Andrew Baggaley for very useful discussions. S.~A. acknowledges financial support from the UK EPSRC (EP/W015730/1), V.~G. from UKRI through a Future Leaders Fellowship (grant number MR/Y018257/1).

We thank ECT* for support at the Workshop ``Nonequilibrium phenomena in superfluid systems: atomic nuclei, liquid helium, ultracold gases, and neutron stars'' during which this work has been developed.

This article is based upon work from COST Action CA24139, Superfluid Condensates in Astrophysics and Laboratory Experiments (SCALES), supported by COST (European Cooperation in Science and Technology).

The data that support the findings of this article are openly available \cite{data_container}. 

\end{acknowledgments}

%%%%%%%%%%%%%%%%%%%%%%%%%%%%%%%%%%%%%%%%%%%%%%%%%%%%%%%%%%%%%%%%%%%%%%%%%%%%%%
%%%%%%%%%%%%%%%%%%%%%%%%%%%%%%%%%%%%%%%%%%%%%%%%%%%%%%%%%%%%%%%%%%%%%%%%%%%%%%

%\section*{Supplementary material}

\end{document}